\documentclass[useAMS,usenatbib,fleqn]{mnras}

\usepackage{psfrag,amsmath,amssymb,color,epsfig,rotating}
\usepackage{graphicx}
\usepackage{graphics}
\usepackage{multirow}
\usepackage{tabularx}
\usepackage{geometry}
\usepackage{array}
\usepackage{hyperref}
\usepackage{subfigure}
\usepackage{amsmath}
\usepackage{comment}

\usepackage[T1]{fontenc}
\usepackage{ae,aecompl}

\usepackage[dvipsnames]{xcolor}
\def\HI{{\rm HI}\,}

\def\kpe{k_{\perp}}
\def\kpa{k_{\parallel}}

\title [RSD due to the motions of {\rm HI}  within galaxies]{Redshift space distortions of the 
	{\rm HI} 21-cm intensity mapping signal due to the internal motions within galaxies}  

\author[Sarkar et al.]{Debanjan
	Sarkar$^{1}$\thanks{debanjan@cts.iitkgp.ac.in},Somnath Bharadwaj$^{1,2}$\thanks{somnath@phy.iitkgp.ac.in}\\ $^1$Centre for Theoretical Studies, Indian
	Institute of Technology Kharagpur, Kharagpur - 721302, India \\ $^2$Department of Physics, Indian Institute of Technology Kharagpur,
	Kharagpur - 721302, India}

\date{\today}
\pubyear{2019}

\begin{document}
	\label{firstpage}
	\pagerange{\pageref{firstpage}--\pageref{lastpage}}
	\maketitle

	\begin{abstract}
		The {\rm HI} 21-cm intensity mapping signal experiences redshift space distortions 
		due to the motion of the galaxies which  contain the {\rm HI} as well as the motions 
		of the {\rm HI} gas within the galaxies. A detailed modelling is essential if this 
		signal is to be used for precision cosmology. Considering dark matter only 
		simulations where the {\rm HI} is assumed to reside in galaxies which are associated 
		with  haloes, in this work we introduce a technique to incorporate 
		the {\rm HI} motions within the galaxies. This is achieved through 
		a line profile which accounts for  both the rotational and random 
		(thermal and turbulent) motions of the {\rm HI} within galaxies. The functional 
		form of the double horned line profiles used here is motivated by 
		observations of $z=0$ spiral galaxies. Analyzing the simulated 
		21-cm power spectrum over the redshift range $1 \le z \le 6$ 
		we find that the {\rm HI} motions within galaxies makes a significant 
		contribution that is manifested as an enhancement in the 
		Finger of God (FoG) effect which can be  modelled reasonably well 
		through a Lorentzian damping profile with a single free parameter $\sigma_p$. 
		The value of $\sigma_p$ is significantly enhanced if motions within 
		the galaxies are included. This is particularly important at $z>3$ where 
		$\sigma_p$ is dominated by the internal motions and a measurement of 
		the FoG effect here could provide a handle on the line profiles of 
		high redshift galaxies.
	\end{abstract}

	\begin{keywords}
		methods: statistical, cosmology: theory, large scale structures, diffuse radiation
	\end{keywords}

	\section{Introduction}
	\label{sec:introduction}
	
	Observations of quasar (QSO) absorption spectra suggest
	that the diffuse Inter Galactic Medium (IGM) is highly ionized
	in the post-reionization era ($z \leqslant 6$) 
	\citep{becker-fan01, fan-carilli-keating06, fan-strauss06}.
	The majority of the residual neutral Hydrogen (\HI)
	is hosted by discrete clouds with high \HI
	column density, $N_{\HI}\geqslant 2\times 10^{20}\,{\rm atoms}\,{\rm cm}^{-2}$.
	These discrete clouds are identified as the Damped Lyman$-\alpha$ systems (DLAs)
	in the QSO absorption spectra 
	\citep{storrie-lomb00, wolfe05, prochaska-herbertfort-wolfe05, zafar-peroux-popping-milliard13}. 
	The collective redshifted 21-cm emission 
	from these individual \HI clouds  appears as a diffuse background
	radiation below $1420\,{\rm MHz}$. A statistical detection of the intensity 
	fluctuations in this diffused background
	provides a distinct way of probing the formation and evolution of the 
	large-scale structures (LSS) in the Universe \citep{bharadwaj-nath-sethi01}. 
	This technique, known as the 21-cm intensity mapping, makes it possible to 
	survey large volumes of space using current and upcoming radio telescopes \citep{bharadwaj-sethi01,bharadwaj-pandey03,wyithe-loeb08-fluctuations-in-21cm}. 
	In the post-reionization era, the 21-cm signal remains largely unaffected by 
	the reionization processes and this makes the 21-cm power spectrum a direct tracer of the
	underlying matter power spectrum \citep{wyithe-loeb09}.
	A detection of the Baryon Acoustic Oscillations (BAO) in the 21-cm power
	spectrum holds the possibility of placing tight constraints on the dark energy
	equation of state 
	\citep{chang-pen-peterson08, wyithe-loeb-geil08, masui-mcdonald-pen10, seo-dodelson10}. 
	A precise measurement of the 21-cm power spectrum can also be used to constrain the 
	cosmological parameters independent of the BAO 
	\citep{loeb-wyithe08, bharadwaj-sethi-saini09, obuljen-castorina17}. 
	The measurement of the 21-cm bispectrum can quantify the non-Gaussian features 
	in the large-scale matter distribution \citep{ali06-21cmbispec,guhasarkar13-21cmbispec}. 
	The cross-correlation of the 21-cm signal with a variety of other tracers of the LSS
	like the Lyman-$\alpha$ forest 
	\citep{guhasarkar13-lya-21,carucci17-lya21,sarkar18-HI-lya}, the Lyman-break galaxies 
	\citep{navarro15-LBG21}, weak lensing \citep{guhasarkar10-weaklens}, the 
	integrated Sachs Wolfe effect \citep{guhasarkar09-isw21} {\it etc.}
	have also been proposed 
	to be potential probes of the post-reionization era.

	Efforts have been made to detect the cosmological 21-cm 
	signal at low redshifts ($z\lesssim1$). 
	\citet{lah-chengalur-briggs-colless07,lah-pracy-chengalur-briggs09}
	have used the 21-cm signal
	stacking technique for the Giant Metrewave Radio Telescope (GMRT)
	observations at $z\sim0.4$
	to estimate the average \HI mass of galaxies. 
	Signal stacking has also been used to determine the \HI
	mass density ($\Omega_{\HI}$)
	at $z<0.13$ using the Parkes radio telescope observations 
	\citep{delhaize13-stack},
	and at $z \sim 0.1$ and $z \sim 0.2$
	using the Westerbork Synthesis Radio Telescope observations 
	\citep{rhee13-stack}. 
	\citet{kanekar-sethi-dwarkanath16} have
	also used the signal stacking technique for GMRT observations 
	at $z\sim1.3$ to obtain
	an upper limit on the average \HI 21-cm flux density.
	The first detection of the cross correlation between 21-cm maps 
	and galaxy surveys  at $z\approx0.8$
	was reported in \citet{chang-pen-bandura10}, based on the data from 
	the Green Bank Telescope (GBT) and the DEEP2 optical galaxy redshift survey.
	\citet{masui-switzer-banavar13} have also measured the cross-correlation 
	between the 21-cm maps acquired at GBT and LSS traced 
	by the galaxies in the WiggleZ Dark Energy Survey at $z\approx0.8\,$.
	On the other hand, \citet{switzer-masui-bandura-calin13} have measured the 
	auto-power spectrum of the 21-cm intensity fluctuations using the GBT
	observations and constrained the \HI fluctuations at $z\approx0.8\,$.
	\citet{ghosh-bharadwaj-ali-chengalur11b,ghosh-bharadwaj-ali-chengalur11a} have 
	analyzed $610\,{\rm MHz}$ GMRT observations towards detecting the 21-cm signal
	from $z=1.32\,$. They have subtracted the foregrounds from the observed data 
	and found that the residual signal is consistent with noise. They have used this
	to place an upper bound on the amplitude of the \HI signal.

	Several low frequency instruments are planned to detect the 21-cm 
	signal from the post-reionization era.
	A number of intensity mapping experiments like 
	BINGO\footnote{http://www.jb.man.ac.uk/research/BINGO/},
	CHIME\footnote{https://chime-experiment.ca/}, the Tianlai
	Project\footnote{http://tianlai.bao.ac.cn/},
	GBT-HIM\footnote{http://greenbankobservatory.org/},
	FAST\footnote{http://fast.bao.ac.cn/en/},
	ASKAP\footnote{http://www.atnf.csiro.au/projects/askap/index.html},
	HIRAX\footnote{https://www.acru.ukzn.ac.za/hirax/}
	have been planned to survey the intermediate redshift range ($z\sim0.5-2.5$),
	where their primary goal is to detect the BAO.
	The linear radio-interferometric array
	OWFA\footnote{http://www.ncra.tifr.res.in/ncra/research/research-at-ncra-tifr/research-areas/ort/ORT}
	aims to measure the 21-cm fluctuations at $z\sim3.35\,$.
	On the other hand, instruments like the Upgraded GMRT\footnote{http://www.gmrt.ncra.tifr.res.in}
	and SKA\footnote{https://www.skatelescope.org/} hold the potential to
	cover a large redshift range.

	The 21-cm signal is inherently very weak, and it is important 
	to correctly model the expected signal in order to make robust predictions
	for the detectability of the signal with the various instruments
	({\it e.g.} \citealt{bull-ferreira15-late-time-cosmology-with-21cm-IM, sarkar-bharadwaj-ali17-OWFA}).
	Accurate modelling of the 21-cm signal is also necessary to 
	correctly interpret the detected 21-cm power spectrum and
	precisely infer the	cosmological model parameters 
	\citep{obuljen-castorina17, hamsa-refregier19}. 
	A careful modelling of the 21-cm power spectrum 
	can also constrain high redshift astrophysics \citep{IM-report2017}. 
	Modelling is also required as input to validate foreground removal 
	and avoidance techniques ({\it e.g.} see \citealt{choudhuri17-thesis-FG} and
	references therein).

	Considerable efforts have been made towards modelling the
	expected redshifted \HI 21-cm signal. \citet{marin-gnedin10} have used an analytic 
	prescription for assigning \HI to haloes and modelled the bias parameter for the
	\HI selected galaxies. In an alternative approach, \citet{bagla10} 
	have used several analytical prescriptions to populate \HI in the haloes,
	identified from the dark matter only simulations, to model the \HI distribution.
	The same approach has also been used by \citet{khandai-sethi-dimatteo11} and 
	\citet{tapomoy-mitra-majumder12} to study the \HI power spectrum and bias.
	\citet{villaescusa-navarro-viel-datta-choudhury14} have used similar
	analytic prescriptions in conjunction with smoothed particle hydrodynamics
	(SPH) simulations to study the \HI distribution. A number of 
	cosmological hydrodynamical simulations have also been used to investigate 
	the \HI content of the post-reionization universe 
	\citep{dave-katz-openheimer13, barnes-haehnelt-14, villaescusa-navarro16-HI-in-galaxy-clusters}.
	\citet{hamsa-refregier17-halomodel1} have developed a halo model based
	analytical framework that describes the distribution and evolution of \HI across redshifts. 
	\citet{hamsa-refregier-amara17-halomodel2} have used the same formalism
	to constrain the abundance and clustering of the \HI systems at $z<6\,$. 
	\citet{castorina-villaescusa-navarro16} have also developed an independent halo model
	based formalism to describe the \HI clustering. Higher order perturbation
	theory based methods have also been used to study the non-linear effects
	on the \HI power spectrum and bias
	\citep{umeh-maartens-santosh16, umeh17, aurelie-umeh-santos17-HIbias}.   
	\citet{hamsa-girish17} have employed the abundance matching
	technique to quantify the observational constraints on the
	\HI mass - halo mass relation in the post-reionization era.

	In \citet{sarkar-bharadwaj-2016} (hereafter Paper I), we have used an
	analytic prescription, originally proposed by \citet{bagla10}, to populate
	the simulated dark matter haloes with \HI and model the  
	\HI distribution in the redshift range $1 \leqslant z \leqslant 6\,$.
	We have also quantified the evolution of the \HI bias across this $z$ range 
	for $k$ values lying in the range $0.04 \leqslant k/{\rm Mpc}^{-1} \leqslant10\,$.
	Paper I and most of the other works discussed above are, however,
	limited by the fact that they do not consider the redshift 
	space distortion (RSD) \citep{kaiser87} introduced
	by the peculiar motions, which plays a crucial 
	role in the 21-cm intensity mapping \citep{bharadwaj-ali04}.

	The RSD in galaxy redshift surveys is a very important field 
	of study (see \citealt{hamilton-98-rsd-review,lahav04-rsd-review} for reviews).
	The individual galaxies are the fundamental elements in galaxy redshift
	surveys. The surveys are only sensitive to the peculiar motion of the
	individual galaxies and motions inside the galaxies are of no
	consequence for redshift space distortions. 
	In contrast, 21-cm intensity mapping does not see the
	individual sources but only sees the collective radiation from all the
	sources. The signal is sensitive to all motions, be it that of the galaxies
	which contain the \HI or the motion of the \HI inside the galaxies.
	The intensity mapping experiments also have high enough 
	frequency resolution ($< 100$ kHz or $< 20\,{\rm km\,s}^{-1}$) to resolve
	the velocity structure of \HI within the galaxies and 
	the RSD effect here is quite different from that in galaxy surveys.

	In \citet{sarkar-bharadwaj-2018} (hereafter Paper II), we have modelled 
	and predicted the 21-cm signal in redshift space (as against real space).
	Building up on Paper I where we have used a mass assignment scheme to
	populate the haloes with \HI, we have used two separate methods to
	incorporate the RSD due to the peculiar motion of the \HI.  
	In the first method, we have placed the total \HI of a halo at 
	the halo centre of mass and have assumed that the \HI moves with 
	the mean velocity of the host halo.
	We have labelled this as the Halo Centre (HC) method. This method ignores the
	motion of \HI inside the haloes. 
	In the second method, we have distributed the total 
	\HI content of a halo equally among all the dark matter particles
	that constitute the halo and have assumed that the \HI moves with 
	the same velocities as the host dark matter particles.
	We refer to this as the Halo Particle (HP) method. The real
	space clustering of \HI is same for the two methods,
	and they only differ in the redshift space.
	We have modelled the redshift space \HI power spectrum 
	$P^s_{\HI}(\kpe,\kpa)$ with the assumption that it is a
	product of three terms: $(i)$ the real space \HI power spectrum
	$P_{\HI}(k)=b^2(k) P(k)$, where $P(k)$ is the dark matter
	power spectrum in real space and $b(k)$ is the \HI bias,
	$(ii)$ a Kaiser enhancement \citep{kaiser87} factor,
	and $(iii)$ an independent Finger of God damping
	\citep[FoG;][]{jackson72-FoG} term 
	which has $\sigma_p$ the pair velocity dispersion as a free parameter.
	Considering several possibilities for the FoG damping, we have found that
	the Lorentzian damping profile provides a reasonably good fit to the simulated
	$P^s_{\HI}(\kpe,\kpa)$ over the entire range $1 \leqslant z \leqslant 6\,$. 
	The model predictions for the monopole $P^s_0(k)$
	of $P^s_{\HI}(\kpe,\kpa)$ are  consistent with the simulated $P^s_0(k)$
	for $k < 0.3\,{\rm Mpc}^{-1}$ over the entire $z$ range, however, the same
	is true for a limited $z$ ($z\leq3$) for the quadrupole $P^s_2(k)$. At $z>3$,
	the models underpredict $P^s_2(k)$.

	In a recent work, \citet{navarro-ingredients-for-IM2018} have used
	advanced magneto-hydrodynamic simulations to study the 
	redshift space clustering of \HI in the range $z\leq5$
	and found that the ratio of the monopole of the 
	redshift space \HI power spectrum to the real space \HI power spectrum
	can be modelled up to a large $k$ 
	range using the models described in Paper II.
	In another work, \citet{ando-rsd-2018} have
	used cosmological hydrodynamical simulations to investigate 
	the scale dependence and redshift dependence
	of the 21-cm power spectrum in the range $1 \leq z \leq 5\,$.
	They have modelled the RSD effects on $P^s_{\HI}(\kpe,\kpa)$
	at moderately non-linear scales, and
	concluded that the real space \HI bias can be recovered by modelling  
	$P^s_{\HI}(\kpe,\kpa)$ and the bias will not be affected much by the 
	detailed astrophysical effects.

	The HC method of Paper II assumes that the \HI content of a halo
	moves with the mean peculiar velocity of the halo centre of mass.  
	This ignores the motion of the \HI within the haloes (or associated galaxies). 
	The motion of the \HI gas inside the galaxies is expected to contribute to the
	FoG effect in redshift space. In contrast, the HP method of Paper II distributes
	the \HI content of a halo equally among the associated
	dark matter particles and assumes that the \HI moves 
	with the same velocities as the host dark matter particles. The HP
	method possibly overestimates the motion of \HI gas within the galaxies. 
	\citet{navarro-ingredients-for-IM2018} and \citet{ando-rsd-2018} 
	have carried out high resolution magneto-hydrodynamic simulations which 
	they use to follow the evolution of HI. These simulations, which incorporate 
	a large variety of sub-grid astro-physical processes 
	\citep{vogelsberger13, weinberger17, aoyama17, pillepich18, weinberger18}, 
	have been fine tuned to match a variety of observations which include 
	galaxy stellar mass function at $z=0$,	galaxy and halo sizes, colours, 
	metallicities, magnetic fields, clustering {\it etc}. While these simulations 
	do incorporate the \HI motion inside the galaxies, it is not clear if these motions 
	are consistent with the observations of the  21-cm line profiles 
	({\it e.g.} \citealt{koribalski04-HIPASS,walter-THINGS08}).  
	Further, such high resolution magneto-hydrodynamic simulations are computationally 
	extremely expensive and are restricted to relatively small box sizes. 
	It is therefore worthwhile to consider techniques  that couple inexpensive 
	and easily scalable dark matter only simulations with observationally 
	motivated models for the \HI motions within the galaxies.

	Low redshift observations ({\it e.g.} \citealt{koribalski04-HIPASS, walter-THINGS08}) 
	indicate that the \HI is predominantly
	contained in the discs of rotating spiral galaxies. Considering
	a single galaxy, the motion of the \HI is broadly the sum of
	two contributions arising respectively from the rotation of the disk and
	the random motions. The random motions again have two contributions,
	namely thermal and turbulent \citep{wright1974-HI,binney-tremaine-2ED-2008}.
	Due to these motions of \HI, the 21-cm 
	emission from a galaxy has a span over a finite frequency range which is 
	modelled in terms of the total galaxy line profile. 
	Observations show that the galaxies exhibit a variety of line profiles
	with different shapes 
	\citep{bottinelli80, bottinelli82, haynes97-LP, huchtmeier00-LP, koribalski04-HIPASS, walter-THINGS08}. 
	The shape of the galaxy line profiles is primarily determined by the
	kinematics of the \HI within the galaxies and only secondarily by
	the \HI distribution. The spiral galaxies mainly exhibit double-horned
	line profiles ({\it e.g.} \citealt{walter-THINGS08}). The characteristic
	double-horned signature arises primarily from the rotation of the disk.
	Some galaxies also show single-peaked profiles which result either
	when the galaxy is viewed close to face-on \citep{lewis84-LP-faceon} or, as is commonly
	found in irregular galaxies, when random motions of \HI dominate \citep{begum08-LP-irreg}.

	Previously \citet{bharadwaj-srikant04} and 
	\citet{chatterjee-bharadwaj-marthi-17-OWFA} have 
	studied the role of line profile
	of discrete \HI sources in shaping the statistical properties of 
	the redshifted 21-cm signal.
	For all the discrete \HI sources in their simulations, they have used 
	flat line profiles, which is not related to the \HI mass of the sources.
	They have showed that the \HI 21-cm visibility correlation, 
	at the smaller frequency separations which are comparable to the
	line width of the emission, changes with the change in the width of the line profile.

	In this work, we use a realistic model to incorporate the motions of \HI within
	the galaxies. We assume that each halo hosts a single galaxy that
	moves with the centre of mass velocity of the halo. Each galaxy mainly consists
	of two parts, a core (or bulge) and a surrounding disk, and the galaxy as a whole
	rotates about the centre of mass. The total \HI content of a galaxy is divided
	between the core and the disk. We also assume that the random motion of 
	the \HI in a galaxy can be modelled by a one-dimensional Gaussian distribution.
	We use a realistic model for the total line profile of the galaxy which
	combines both the rotational and the random motion effects. We use this model 
	to simulate line profiles for all the galaxies and map the \HI distribution in
	redshift space. We study the effects of the galaxy line profile on
	$P^s_{\HI}(\kpe,\kpa)$ and try to model the RSD anisotropies 
	using the same models as described in Paper II. In reality, a halo can
	host more than one galaxy, and moreover the galaxies can have their
	own motion inside the haloes. We have not considered these here.

	This paper is organized as follows. We present the real space \HI simulations in
	Section~\ref{subsec:real-HI}. In Section~\ref{subsec:simulating-LP}, we describe the
	technique to simulate the \HI line profiles of the haloes. We present our findings in
	Section~\ref{sec:results}, and Section~\ref{sec:summary-discussion} contains 
	summary and discussion.

	We have adopted the best-fit cosmological parameters from \citet{planck-collaboration15}. 
	
	% ******************************************************* %
	
	\section{Simulating the Redshift Space \HI Distribution}
	\label{sec:method}
	
	% ******************************************************* %
	
	\subsection{Simulating the Spatial \HI Distribution}
	\label{subsec:real-HI}
	
	We first simulate the post-reionization \HI distribution in real space. 
	We have used a Particle Mesh N-body code \citep{bharadwaj-srikant04} to 
	generate snapshots of dark matter distribution in the redshift range $z=1-6$ 
	with redshift interval $\Delta z = 0.5\,$. The simulations include $[1,072]^3$
	dark matter particles in a $[2,144]^3$ regular cubic grid of comoving spacing
	$0.07\,{\rm Mpc}$ with a total simulation volume (comoving) of 
		$[150.08\,{\rm Mpc}]^3$. The grid spacing roughly translates to a mass resolution of 
	$10^8\,{\rm M}_{\odot}$. The simulations used here are the same as those 
	analyzed in Papers I and II.

	We employ the Friends-of-Friends (FoF) \citep{davis85} algorithm, with a 
	linking length of $0.2$ in the unit of the mean inter particle
	separation, to locate the collapsed haloes in the dark matter
	snapshots. We define an assembly of dark matter particles having 
	$10$ or more members as a halo. This sets the halo mass resolution to
	$10^9\,{\rm M}_{\odot}$, which is adequate for the reliable prediction 
	of the 21-cm brightness temperature fluctuations \citep{kim-wyithe-baugh-lagos16}.
	The minimum halo mass identified by the FoF algorithm has a fixed value
	$10^9\,{\rm M}_{\odot}$ at all redshifts. On the other hand, the most 
	massive haloes at redshifts $z=(1,2,3,4,5,6)$ have masses 
	$\approx(320,100,40,18,9,3)\times10^{12}\,{\rm M}_{\odot}$.

	In the post-reionization era ($z \le 6$), the bulk of the \HI is accommodated
	in the highly dense objects with \HI column density 
	$N_{\HI} \geqslant 2 \times 10^{20}\,{\rm atoms\,cm}^{-2}$,
	where they remain protected from the ionizing radiation. These objects can be identified 
	with the broad damping wings observed
	in quasar absorption spectra and these are labelled as the
	Damped Lyman-$\alpha$ Systems (DLAs)
	\citep{storrie-lomb00,prochaska-herbertfort-wolfe05,zafar-peroux-popping-milliard13}.
	DLAs are believed to be associated with galaxies \citep{haehnelt-steinmetz-rauch00} and
	are hosted in the dark matter haloes within the mass range $10^9 < M_h/{\rm M}_{\odot} < 10^{12}$ 
	\citep{cooke-wolfe-gawiser06,pontzen08,font-ribera12-cross-correlation-DLA-Ly-alpha}.

	We have assumed that the \HI is solely contained within the dark matter haloes,
	which is a fairly good assumption at these redshifts 
	\citep{villaescusa-navarro-viel-datta-choudhury14,navarro-ingredients-for-IM2018}.
	We also expect that the \HI mass $M_{\HI}$ of a halo will increase with the host
	halo mass $M_h$. However, in the local Universe, we observe that the massive 
	elliptical galaxies and the galaxy clusters contain negligible amounts of \HI
	\citep{serra-oosterloo-morganti12}. Therefore, there should be a maximum 
	limit in halo mass $M_h$ beyond which the \HI content	diminishes. On the other 
	hand, haloes with mass below a critical lower limit would not be able to self-shield the neutral 
	gas against the harsh radiation and therefore cannot host \HI. Based on the above 
	considerations, \citet{bagla10} have proposed three analytic prescriptions to populate 
	the simulated haloes with \HI. Papers I and II have utilized their third prescription
	(Equation 6 of \citealt{bagla10}) to simulate the post-reionization \HI distribution.
	In this work, we have used their first prescription (Equation 4 of \citealt{bagla10}) 
	to populate the haloes with \HI. \citet{bagla10} have used a redshift-dependent relation 
	between $M_h$ and circular velocity
	$v_{circ}$
	% ----------------------------------
	\begin{equation} 
	M_h = 10^{10} \,{\rm M}_{\odot}\, \Big( \frac{v_{circ}}{60 \, {\rm km}\, {\rm s}^{-1}} \Big)^3 \Big( \frac{1+z}{4} \Big)^{-\frac{3}{2}}\,.
	\label{eq:virial_relation}
	\end{equation} 
	% ----------------------------------
	The prescription assumes that the haloes with a minimum circular velocity 
	$v_{circ} = 30 \, {\rm km}\, {\rm s}^{-1}$ will host \HI which decides the lower mass 
	limit $M_{\rm min}$. The lower limit of circular velocity is consistent with a recent finding by  
	\citet{villaescusa-navarro16-HI-in-galaxy-clusters} which states that the haloes with 
	circular velocities larger than $\sim 25 \, {\rm km}\, {\rm s}^{-1}$ are required to host 
	\HI in order to reproduce observations. The upper mass limit $M_{\rm max}$ is decided by 
	setting $v_{circ} = 200 \, {\rm km}\, {\rm s}^{-1}$. According to the prescription, the 
	\HI mass $M_{\HI}$ in a halo is related to the halo mass $M_h$ as
	% ----------------------------------
	\begin{equation}
	M_{\rm \HI}(M_h) = \left\{  \begin{array}{l l}	 
	f_1 ~ M_h & \quad
	\text{if $M_{\rm min} \leqslant M_h \leqslant M_{\rm max}$} \\ 0 & \quad
	\text{otherwise}\\
	\end{array} \right.  \,,
	\label{eq:bagla_scheme1}
	\end{equation}
	% ----------------------------------
	where $f_1$ is a free parameter that decides the amount of 
	\HI to be put in the simulation volume and it is fixed in such a way that the
	simulated cosmological \HI mass density $\Omega_{\HI}$ remains fixed at 
	a value $\sim 10^{-3}$. However, the results of this work are not
	sensitive to the choice of $f_1$.

	The \HI assignment prescription adopted here assumes that the haloes with masses
	greater than $M_{\rm max}$ do not contain \HI. This reflects the fact that the 
	most massive galaxies in the local Universe,  which are largely ellipticals,   
	contain negligible amounts of \HI	\citep{serra-oosterloo-morganti12}.
	However, it is possible that the more massive haloes with masses  greater 
	than $M_{\rm max}$ contain more than one galaxy, 
	of which some may host significant amounts of \HI. \citet{navarro-ingredients-for-IM2018} 
	provide	alternative \HI assignment prescriptions  by fitting
	state-of-the-art hydrodynamic simulations which leads to a  
	non-negligible \HI   contribution from  halo with masses above $M_{\rm max}$
	\citep{modi19}. This is further confirmed by the recent work of \citet{obuljen19} 
	where the authors model the abundance and clustering of \HI by combining the 
	Arecibo Legacy Fast ALFA survey (ALFALFA) and the Sloan Digital Sky Survey (SDSS) data.
	The simulations also show that for the massive haloes, the satellite galaxies		
	contain a significant fraction of \HI.  In this work  we have only considered 
	a single galaxy per halo and adopted  the simple \HI assignment prescription presented in Equation~\ref{eq:bagla_scheme1}     to avoid the complications 
	arising from having multiple galaxies in a single halo. Allowing for this will 
	lead to an increase in the FoG suppression. In this paper the main focus  is to 
	highlight the RSD contribution  from internal motions within galaxies, and we  
	expect  this effect to be significant  even if we consider alternate 
	\HI prescriptions.

	In our simulations, we cannot resolve the haloes with mass below
	$10^9\,{\rm M}_{\odot}$ and at $z>3.5$, $M_{\rm min}$ falls below our mass resolution.
	We have used $M_{\rm min}=10^9\,{\rm M}_{\odot}$ for $z>3.5$ in the \HI population prescription
	(Equation~\ref{eq:bagla_scheme1}). For $z \leqslant 3.5$, $M_{\rm min}$ stays above our halo
	mass resolution and we can fully resolve the smallest haloes that host \HI. Our earlier work
	(Paper I) shows that the choice of $M_{\rm min}=10^9\,{\rm M}_{\odot}$ for $z>3.5$
	does not affect the results significantly.

	% **************************************************** %
	\subsection{Simulating the \HI line profile of the haloes}
	\label{subsec:simulating-LP}
	%\subsection{The \HI Line Profile}
	%\label{subsec:LP}
	
	% ========================================================
	The intensity mapping signal is observed as a function of frequency (or redshift) 
	and thus it contains the Doppler effect of the \HI motions along the line of sight.
	This causes the \HI distribution observed in redshift space to be rearranged with 
	respect to real space. For a distant observer along the $z$-axis, 
	the position in redshift space ${\mathbf s}$ is related to the position in real space ${\mathbf x}$ by
	% ----------------
	\begin{equation}
	{\mathbf s}={\mathbf x} + \frac{v \hat{z}}{aH(a)}\,,
	\label{eq:mapping}
	\end{equation}
	% ----------------
	where, $v$ is the line of sight component of the peculiar velocity, $a$ is the scale factor and
	$H(a)$ is the Hubble parameter. We assume that each halo hosts a single galaxy which 
	contains the entire \HI mass associated with the halo. We further assume that the galaxy
	as a whole moves with the mean velocity, say $v_h$, of the halo. It is also possible that the 
	galaxy may have a motion within the halo, however, we have not considered this in our simulations.

	We now consider the motion of the \HI within a galaxy. 
	This motion can be categorized broadly into two parts $-$
	ordered motion and random motion. The ordered motion here 
	refers to the rotational motion, and the random motion 
	consists of thermal motion and turbulence. These motions cause the \HI mass 
	distribution in redshift space to appear smeared along the line of sight. 
	We have modelled this in terms of a line profile $\phi(v)$, where the 
	\HI mass element $\Delta M_{\HI} (v)$ moving with line of sight
	velocity in the range $v$ to $v + \Delta v$ is given by,
	% ----------------
	\begin{equation}
	\Delta M_{\HI} (v) = M_{\HI} \times \phi(v) \Delta v\,,
	\label{eq:total_HImass}
	\end{equation}
	% ----------------
	where, $v$ is with respect to the rest frame of the galaxy (or the host halo). 
	Here, $M_{\HI}$ refers to the total \HI mass of the halo and $\int \phi(v) dv=1\,$.

	The observed 21-cm line profile of galaxies have different shapes
	\citep{bottinelli80, bottinelli82, haynes97-LP, huchtmeier00-LP, koribalski04-HIPASS, walter-THINGS08}.
	However, for rotating galaxies, the line profile mainly exhibits a 
	double horned-shape \citep[{\it e.g.} see][]{walter-THINGS08}.
	There have been a number of efforts to model the 21-cm line profile and explain its shape.
	\citet{roberts-21cm-linewidth1978} has discussed the 21-cm emission line profile for rotating galaxies
	and explained why these often exhibit a double-horned shape. \citet{roberts-21cm-linewidth1978} 
	has also presented some semi-physical models in order to explain the observed line
	profiles. \citet{obreschkow-09a,obreschkow-09b} have presented
	multi-parameter models that can describe both \HI and CO emission
	lines. \citet{westmeier14-busyfunction} have proposed a new analytic function, named the `busy function', 
	to accurately describe the double-horned \HI profiles of galaxies. \citet{stewart-HI-profile14} have presented a	physically-motivated six parameter model for the \HI emission profile which
	they have used to fit and explain the profiles of the THINGS galaxies. We have modified the \citet{stewart-HI-profile14} 21-cm line profile model and incorporated this in our simulations. 
	
	%----------------------------------- Figure ---------------------------	
	\begin{figure*}
		
		\psfrag{HI profile}[c][c][1.4][0]{$\phi(v)$}
		\psfrag{velocity}[c][c][1.4][0]{$v\,{\rm km\,s}^{-1}$ }
		
		\psfrag{al0.0}[c][c][1.2][0]{$\,\alpha=\mathbf{0}\qquad\quad$}
		\psfrag{al0.4}[c][c][1.2][0]{$0.4$}
		\psfrag{al0.8}[c][c][1.2][0]{$0.8$}
		\psfrag{al1.0}[c][c][1.2][0]{$1.0$}
		
		\psfrag{hf0.0}[c][c][1.2][0]{$h_f=0\qquad \quad \;\;$}
		\psfrag{hf0.1}[c][c][1.2][0]{$\mathbf{0.1}$}
		\psfrag{hf0.4}[c][c][1.2][0]{$0.4$}
		\psfrag{hf0.8}[c][c][1.2][0]{$0.8$}
		\psfrag{hf1.0}[c][c][1.2][0]{$1.0$}
		
		\psfrag{sigmav5}[c][c][1.2][0]{$\sigma_v=\;\;5\,{\rm km\,s}^{-1} \;\;\;\qquad$}
		\psfrag{sigmav10}[c][c][1.2][0]{$\mathbf{10}\,{\rm km\,s}^{-1}\,$}
		\psfrag{sigmav20}[c][c][1.2][0]{$20\,{\rm km\,s}^{-1}\,$}
		
		\centering
		\includegraphics[width=0.37\textwidth,angle=-90]{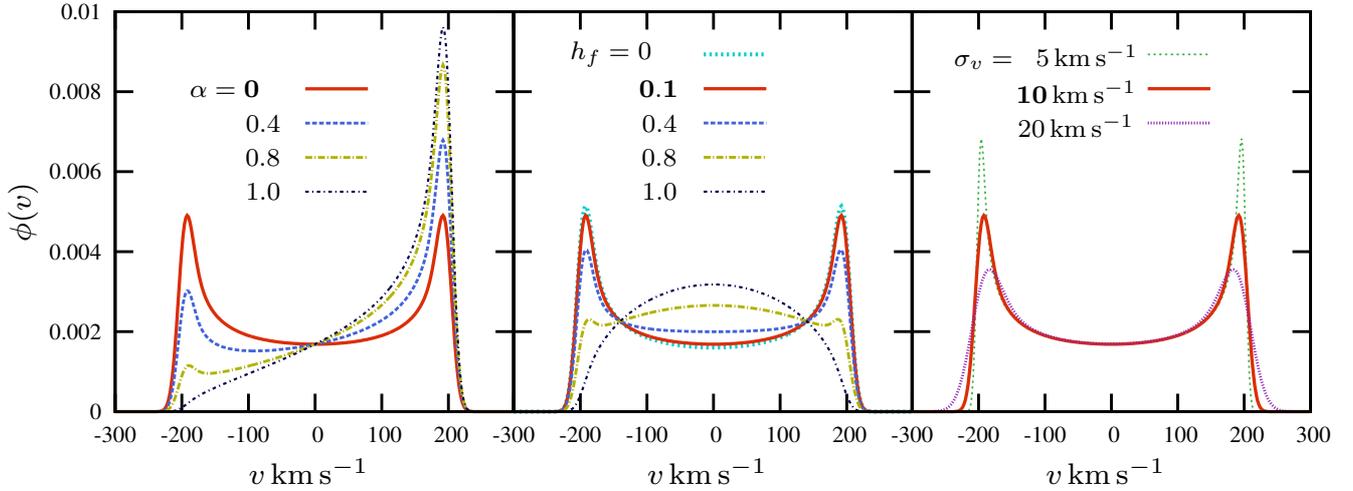}
		
		\caption{The left, central and right panels show how the   
			\HI line profile $\phi(v)$ (Equation~\ref{eq:complete-line-profile}) 
			varies with respect to changes in the 
			model parameters $\alpha$, $h_f$ and $\sigma_v$ respectively. 
			The  reference model with parameter values 
			$\alpha=0, h_f=0.1,\sigma_v=10\,{\rm km\,s}^{-1}$ 
			(red solid curves) is shown in all the panels. Here we have considered
			$v_{circ}=200\,{\rm km\,s}^{-1}$ and $i=0$ for all the line profiles.}
		
		\label{fig:model-profiles}
	\end{figure*}
	%----------------------------------- Figure ---------------------------		

	Following \citet{stewart-HI-profile14} we have 
	modelled the galaxy as having broadly two parts, a core (or bulge) 
	and a gaseous disk encircling the core. 
	The \HI is mainly found in the disk, the core contains a small amount of \HI.
	The core and the disk are both in circular motion around the centre of the galaxy.
	The angular velocities of both of these components lie along a common axis which is 
	perpendicular to the plane of the disk. The circular velocity of a typical galaxy
	starts from zero at the galactic centre, rises steadily with radius and
	then saturates at the disk. We denote this saturated value of the
	circular velocity as $v_{circ}$ which depends on the mass of the
	host halo $M_h$ through Equation~\ref{eq:virial_relation}.

	Note that, the circular velocity of a galaxy may be different from
	the circular velocity of its host halo (Equation~\ref{eq:virial_relation})
	and the two quantities can be related through a factor $f_v$ 
	\citep{mcgaugh12-BTFR} which can vary with the mass (or the circular velocity)
	of the disk galaxies (e.g., \citealt{bullock01-BTFR}). Studies show that 
	for the massive disk galaxies with circular velocity $> 100\, {\rm km\,s}^{-1}$, 
	$f_v$ lies in the range $1 \le f_v \le 1.3$ \citep{sellwood05-BTFR,reyes12}.
	On the other hand, $f_v \lesssim 1$ for the lower mass galaxies \citep{stark09-BTFR}.
	However, the estimations of the factor $f_v$ are limited only to very 
	low redshifts ($z<1$). We do not have much knowledge about $f_v$ at high 
	redshifts. In order to keep our analysis simple, we assume $f_v=1$
	which implies that a galaxy and its host halo have 
	same circular velocities.

	Assuming an inclination angle `$i$' between the observer's line of sight
	and the normal to the galaxy's disk, the observed circular
	velocity is $v_c = v_{circ} \times \sin(i)$.
	The rotation of the disk mainly gives rise to the characteristic 
	double-horned profile that has a 
	central minimum and two maxima around $\pm v_c$. 
	Following \citet{stewart-HI-profile14}, 
	we model the rotating core as a half ellipse. Combining the core and the disk, 
	the total line profile for the rotational motion of a galaxy can be modelled as  
	% ----------------------------
	% \begin{eqnarray}
	\begin{equation}
	\begin{aligned}
	S(v)  = & \mathcal{A}\left(1+ \frac{\alpha \, v}{v_{circ}}\right) \Big[ (1-h_f)/\sqrt{1-(v/v_c)^2} \\
	+ & h_f \sqrt{1-(v/v_c)^2} \Big]  \,,{\rm with} \, \mid \frac{v}{v_c} \mid <1 \,.
	\end{aligned}
	\label{eq:bulk-profile}
	\end{equation}
	% \end{eqnarray}
	% ----------------------------
	Here, $\mathcal{A}$ is an arbitrary normalization factor, the first term
	in the square bracket corresponds to the disk that gives rise to the
	double-horned feature, the second term in the square bracket corresponds 
	to the core, $h_f$ is the fraction of \HI found in the core. The natural
	range of $h_f$ is $0 \leqslant h_f \leqslant 1\,$. An asymmetry in the line 
	profile can be accommodated by the factor $\left(1+ \frac{\alpha \, v}{v_{circ}}\right)$ 
	where $\alpha$ is the asymmetry parameter which varies between 
	$-1 \leqslant \alpha \leqslant 1\,$.
	For convenience, we have only used the values $0\leqslant \alpha \leqslant 1$, as
	the negative values would give similar results.
	Considering the random motions, we assume
	that the projection of the random velocities along the 
	line of sight has a one-dimensional Gaussian distribution,
	% ----------------------------
	\begin{equation}
	G(v)=1/\sqrt{2 \pi \sigma^2_v} \exp (-v^2/2 \sigma^2_v)\,,
	\label{eq:random-profile}
	\end{equation} 
	% ----------------------------
	where $\sigma_v$ is the velocity dispersion of \HI in the Inter Stellar Medium (ISM). 
	For the nearby galaxies, the typical values
	of $\sigma_v$ are close to $\sim 10 \,{\rm km\,s}^{-1}$ 
	({\it e.g.} see \citealt{petric-rupen07} and references therein).
	The complete line profile is the convolution
	of the rotational and the random profiles,
	% ----------------------------
	\begin{equation}
	\phi(v)=\int S(u)G(v-u) du \,.
	\label{eq:complete-line-profile}
	\end{equation}
	% ----------------------------
	Figure~\ref{fig:model-profiles} shows the model \HI line profiles for
	different values of $\alpha$, $h_f$ and $\sigma_v$.

	%----------------------------------- Figure ---------------------------			
	\begin{figure}
		
		\psfrag{phi}[c][c][1.5][0]{$\phi(v)$}
		\psfrag{v}[c][c][1.5][0]{$v$}
		\psfrag{vk}[c][c][1.3][0]{\textcolor{red}{$v_k$} }
		\psfrag{deltav}[c][c][1.6][0]{$\Delta v$}
		\psfrag{HI mass of a particle}[c][c][1.1][0]{$\qquad \qquad \Delta M_{\HI}(v_k) = M_{\HI} \phi(v_k) \Delta v $}
		
		\centering
		\includegraphics[width=.43\textwidth,angle=0]{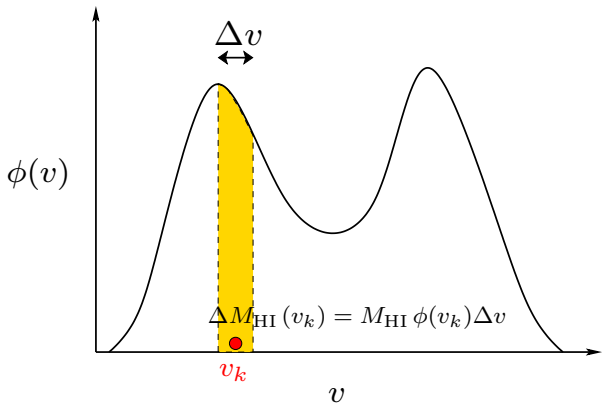}

		\caption{This figure depicts the procedure of dividing the total line profile $\phi(v)$
			of a galaxy into discrete \HI mass elements. The line profile $\phi(v)$ 
			is first divided into a number of velocity bins	of equal width $\Delta v$. 
			The shaded region here represents the area under the $k^\text{th}$ 
			bin with velocity in the range $v_k$ to $v_k+\Delta v$. 
			The velocities are with respect to the rest frame of the host halo.
			The \HI mass enclosed under the $k^\text{th}$ bin is 
			$\Delta M_{\HI}(v_k) = M_{\HI} \phi(v_k) \Delta v$, where $M_{\HI}$ 
			is the total \HI mass of the halo. We consider this mass element 
			$\Delta M_{\HI}(v_k)$ as a particle which moves with a velocity $v_k$. 
			The total \HI mass $M_{\HI}$ is the sum of the all mass elements because 
			the line profile here is normalised to $\int \phi(v) dv=1\,$.}
		
		\label{fig:discrete-HI}
	\end{figure}
	%----------------------------------- Figure ---------------------------		

	The real space simulations discussed in Section~\ref{subsec:real-HI} provide us
	with halo catalogues containing $(i)$ the halo mass $M_h$, 
	$(ii)$ halo position and $(iii)$ mean velocity $v_h$ of the halo.
	Following the \HI assignment scheme (Equation~\ref{eq:bagla_scheme1}), we obtain the
	\HI content $M_{\HI}$ of all the haloes. We assume that each halo hosts a
	disk galaxy at the halo centre of mass, and that galaxy 
	contains the entire \HI mass and moves with the mean velocity $v_h$
	of the associated halo. Depending upon the host halo mass $M_h$,
	we assign $v_{circ}$ to each such galaxy using Equation~\ref{eq:virial_relation}.
	Here, $v_{circ}$ for any galaxy is defined with respect to the rest frame
	of its host halo. The orientation of the galaxy disks are 
	taken to be random. Considering a single galaxy, the total line profile
	is generated using the model described in Equation~\ref{eq:complete-line-profile}. 
	For simplicity, it is assumed that
	all the galaxies in a simulation have same $\alpha$, $h_f$ and $\sigma_v$.
	As described in Figure~\ref{fig:discrete-HI}, for each galaxy, we divide 
	the total line profile into velocity bins of equal width
	$\Delta v$ and treat the \HI enclosed within individual 
	bins as separate particles. The total velocity of any such 
	particle, with respect to the local rest frame, is the sum of $(i)$ the velocity of
	the respective bin ({\it e.g.} $v_k$ for the $k^\text{th}$ bin)
	and $(ii)$ the line of sight component of  $v_h$.
	We use sum of these two velocities to 
	map the position of each \HI particle
	into redshift space using  Equation~\ref{eq:mapping}. 
	For our analysis, we have considered $\Delta v= 1 \,{\rm km\,s}^{-1}$.

	The observations of the characteristic double-horned line profiles
	\citep{stewart-HI-profile14}
	are all restricted  to  very low redshifts ($z<0.5$). Our understanding of 
	the kinematics of \HI  at high redshifts ($z>0.5$) majorly depends on the 
	study of the  QSO absorption spectra where the absorption features are 
	mainly dominated by the Lyman$-\alpha$ systems \citep{zafar-peroux-popping-milliard13}. 
	The systems with the largest column densities, the  DLAs,  are proposed 
	to be the progenitors of the present day spiral galaxies \citep{wolfe05}. 
	Modelling of DLAs observations suggests that DLAs resemble rotating disk galaxies with 
	circular velocities typically of the order of $100-200\,{\rm km\,s}^{-1}$  
	\citep{wolfe86, wolfe95, lanzetta-wolfe95, jedamzik98, kauffmann94, klypin95}. 
	A number of theoretical studies indicate that the DLAs can have circular velocities 
	as low as $\sim 50 \,{\rm km\,s}^{-1}$ ({\it e.g.} \citealt{kauffmann96}) 
	which is supported by a number of numerical simulations
	\citep{pontzen08, cen12-nature-of-DLAs-and-their-hosts, bird14, bird15}. 
	QSO absorption spectra also suggest that the gas inside the DLAs has a velocity 
	dispersion of $\sigma_v \approx 5-10\,{\rm km\,s}^{-1}$ \citep{wolfe05} which is  
	comparable with the nearby galaxies. Other observations also reveal that the high 
	redshift galaxies show rotational motion ({\it e.g.} see 
	\citealt{pettini09} and references therein). 
	Based on these considerations we have assumed  that  across the entire  range 
	$z \le 6$ the \HI resides in rotating disk galaxies which exhibit a  double-horned 
	line profile similar to those seen for local galaxies. The value of the parameters 
	$\alpha, h_f,\sigma_v$  are found to vary from galaxy to galaxy in the local Universe. 
	The statistics of these parameters and their redshift evolution are largely unknown. 
	In our analysis each simulation corresponds to  fixed values of these parameters which 
	are held constant over the entire $z$ range. We have carried out simulations covering 
	the entire range of parameter values in order to estimate how this variation affects the 
    RSD. Considering  $v_{circ}$ which determines the overall width of the line profile, 
    it 	may be noted that the values are determined by the halo mass distribution which  
    evolves	with $z$.

	The simulation technique outlined above incorporates  the effect of galaxy  line profiles 
	on the redshift space \HI distribution and we refer to this as
	the `LP' method. We use the cloud in cell interpolation to calculate
	the \HI density on the grids and Fourier transform these 
	to calculate the \HI power spectrum. We run five independent 
	realizations of the simulation at every redshift to calculate
	the mean of the power spectrum presented here. 
	
	% **************************************************** %

	\section{Results}
	\label{sec:results}
	
	%----------------------------------- Figure ---------------------------		
	\begin{figure*}
		
		\psfrag{z=1.0}[c][c][1.2][0]{$\;\; z=1$}
		\psfrag{z=2.0}[c][c][1.2][0]{$z=2$}
		\psfrag{z=3.0}[c][c][1.2][0]{$z=3$}
		\psfrag{z=4.0}[c][c][1.2][0]{$z=4$}
		\psfrag{z=5.0}[c][c][1.2][0]{$z=5$}
		\psfrag{z=6.0}[c][c][1.2][0]{$z=6$}
		
		\psfrag{kpar}[c][c][1.3][0]{$\kpa\,{\rm Mpc}^{-1}$}
		\psfrag{kperp}[c][c][1.3][0]{$\kpe\,{\rm Mpc}^{-1}$}
		\psfrag{HC}[c][c][1][0]{$\,$HC}
		\psfrag{HP}[c][c][1][0]{HP}
		\psfrag{Real Space}[c][c][1][0]{Real Space$\quad$}
		\psfrag{HI-vc}[c][c][1][0]{$\quad$LP}

		\centering
		\includegraphics[width=.68\textwidth,angle=-90]{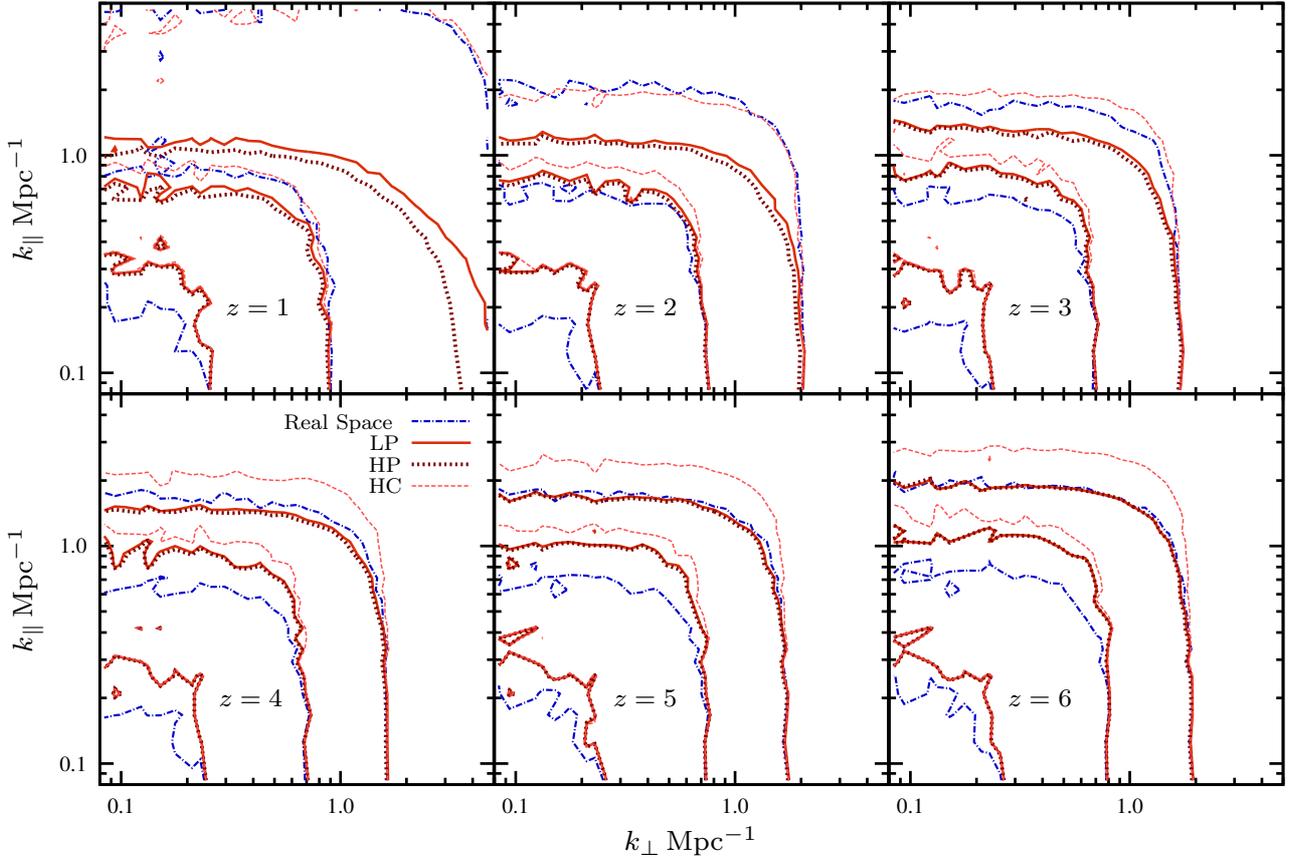}	
		
		\caption{The red contours with different line styles 
			show the redshift space \HI power spectrum 
			$P^s_{\HI}(\kpe,\kpa)$ for the LP, the HC and the HP methods
			at six different redshifts.	The blue contours show the 
			isotropic real space power spectrum $P_{\HI}(k)$.
			For the LP method, we have considered
			the reference values $\alpha=0$, $h_f=0.1$ and 
			$\sigma_v=10\,{\rm km\,s}^{-1}$.}
		\label{fig:2dpowerspectrum}
	\end{figure*}
	%----------------------------------- Figure ---------------------------		

	We have generated the \HI distributions in redshift space using the LP method 
	assuming that all the galaxies in a simulation  have the same values of 
	$\alpha$, $h_f$ and $\sigma_v$.	Throughout this paper we have used the parameter
	values $\alpha=0$, $h_f=0.1$ and $\sigma_v=10\, {\rm km\,s}^{-1}$ as the reference for the
	LP method and we show this with solid red lines in all the figures.
	Figure~\ref{fig:2dpowerspectrum} shows the simulated redshift space \HI power	
	spectra $P^s_{\HI}(\kpe,\kpa)$ at six different redshifts considering the LP method. 
	For comparison, we have  shown $P^s_{\HI}(\kpe,\kpa)$ 
	for the HC and the HP methods introduced in Paper II and 
	summarised in Section~\ref{sec:introduction} of this paper.
	The real space \HI power spectrum $P_{\HI}(k)$, which is isotropic
	with respect to $\kpa$ and $\kpe$, is also shown as a reference for the other models.
	The outer contours correspond to smaller 
	spatial scales (or larger k), and the inner contours correspond to 
	larger spatial scales (or smaller k). The contour values increase inwards.
	Any deviation between $P_{\HI}(k)$ and $P^s_{\HI}(\kpe,\kpa)$ is due to
	redshift space distortion (RSD). There are broadly two types of RSD 
	anisotropies visible here, $(i)$ elongation of $P^s_{\HI}(\kpe,\kpa)$ contours along 
	$\kpa$ due to the Kaiser effect \citep{kaiser87} and $(ii)$ compression of 
	$P^s_{\HI}(\kpe,\kpa)$ contours along $\kpa$ caused by the Finger of God (FoG) effect
	\citep{jackson72-FoG}. 
	At all redshifts, for small $k$ the $P^s_{\HI}(\kpe,\kpa)$ contours for the LP, the HC and the HP methods all overlap. These contours are elongated along  $\kpa$ with respect to $P_{\HI}(k)$  indicating that the three methods predict the same coherent flows (and Kaiser effect) at large length-scales.  At redshifts $z<5$,  for large $k$  
	the $P^s_{\HI}(\kpe,\kpa)$ contours  are compressed along  $\kpa$  with respect to $P_{\HI}(k)$  
	for the LP and the HP methods whereas this compression is only seen at $z <3$ for the HC method. 
	Considering the entire redshift range,  we see that at large $k$ the FoG compression  of the HC method are 
	considerably different from those of the LP and the HP methods.

	%----------------------------------- Figure ---------------------------		
	\begin{figure*}
		
		\psfrag{al=0.0}[c][c][1][0]{$\,\alpha=0\qquad\quad$}
		\psfrag{al=0.4}[c][c][1][0]{$0.4$}
		\psfrag{al=0.8}[c][c][1][0]{$0.8$}
		\psfrag{al=1.0}[c][c][1][0]{$1.0$}
		
		\psfrag{hf=0.1}[c][c][1][0]{$h_f=0.1\qquad\,\:$}
		\psfrag{hf=0.4}[c][c][1][0]{$0.4$}
		\psfrag{hf=0.8}[c][c][1][0]{$0.8$}
		\psfrag{hf=1.0}[c][c][1][0]{$1.0$}
		
		\psfrag{sv=10.0}[c][c][1][0]{$\sigma_v=10 \,{\rm km\,s}^{-1}\qquad\quad$}
		\psfrag{sv=5.0}[c][c][1][0]{$5 \,{\rm km\,s}^{-1}\,\,\,$}
		\psfrag{sv=15.0}[c][c][1][0]{$15 \,{\rm km\,s}^{-1}\,\,\,$}
		\psfrag{sv=20.0}[c][c][1][0]{$20 \,{\rm km\,s}^{-1}\,\,\,$}
		
		\psfrag{hf=0.1 sv=10.0}[c][c][1][0]{$h_f=0.1,\,\sigma_v=10 \,{\rm km\,s}^{-1}\quad$}
		\psfrag{al=0.0 sv=10.0}[c][c][1][0]{$\alpha=0,\,\sigma_v=10 \,{\rm km\,s}^{-1}\quad$}
		\psfrag{al=0.0 hf=0.1}[c][c][1][0]{$\alpha=0,\,h_f=0.1$}
		
		\psfrag{HP}[c][c][1][0]{HP}
		\psfrag{HC}[c][c][1][0]{HC}
		\psfrag{HI monopole}[c][c][1][0]{$P^s_{0}/[P^s_0]_{\rm HC}$}
		\psfrag{kmode}[c][c][1][0]{$k\, {\rm Mpc}^{-1}$}
		
		\psfrag{z=1}[c][c][1.3][0]{$z=1$}
		\psfrag{z=3}[c][c][1.3][0]{$z=3$}
		\psfrag{z=5}[c][c][1.3][0]{$z=5$}
		
		\centering
		\includegraphics[width=0.5\textwidth,angle=-90]{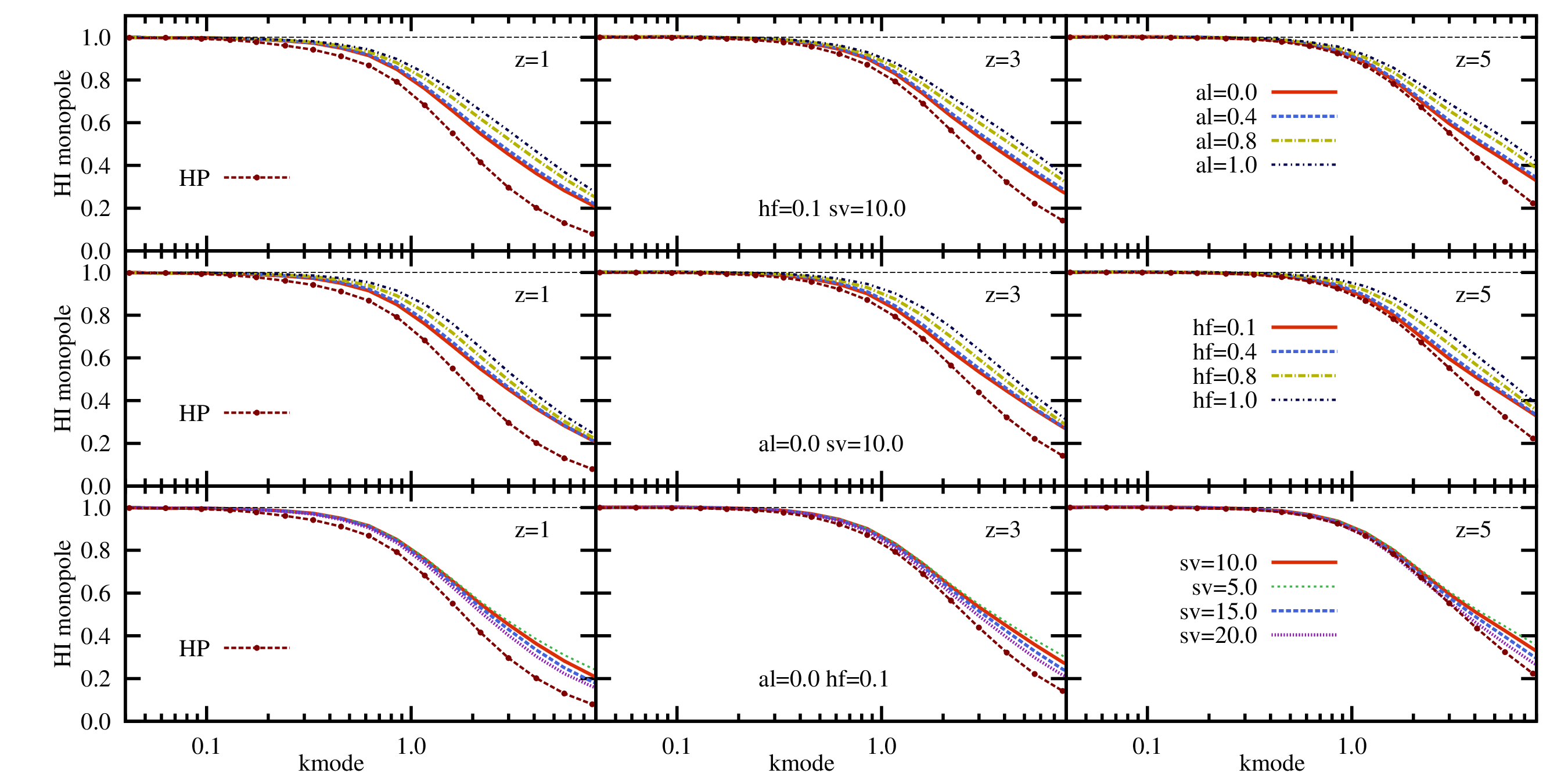}
		
		\caption{This shows the variation of the monopole $P^s_0(k)$ with respect to 
			the parameters $\alpha$ (upper panels), $h_f$ (middle panels) 
			and $\sigma_v$ (lower panels) of the LP method at three different redshifts.
			The variations in $\alpha$, $h_f$ and $\sigma_v$ are with respect to the reference
			values $\alpha=0, h_f=0.1, \sigma_v=10 \,{\rm km\,s}^{-1}$. The results for 
			the HC and the HP methods are also shown for comparison, and 
			all the results have been normalized using $[P^s_0(k)]_{\rm HC}$ of the HC method.}
		
		\label{fig:sim-monopole}
		
	\end{figure*}
	%----------------------------------- Figure ---------------------------		

	The anisotropic \HI power spectrum $P^s_{\HI}(\kpe,\kpa)$ can be decomposed
	into angular multipoles 
	\citep{hamilton-92-RSDcorrelation,cole-fisher-weinberg94}
	% --------------------------
	\begin{equation}
	P^s_{\HI}(\kpe,\kpa)=\sum\limits_{\ell} \mathcal{L}_{\ell}(\mu) P^s_{\ell}(k)
	\end{equation}
	% -------------------------- 
	where $\mu= \kpa/k$ is the cosine of the angle between the line of sight and $\mathbf{k}$ 
	vector, $\mathcal{L}_{\ell}$ are the Legendre polynomials and $P^s_{\ell}(k)$ are 
	different angular moments of the power spectrum. Under the distant observer 
	approximation, only the even angular moments are non-zero. The angular moments 
	are functions of a single variable $k$ and therefore are easy to visualize
	and interpret. From here onwards we shall discuss our results in terms of the
	first two even moments, monopole $P^s_0(k)$ and quadrupole $P^s_2(k)$.
	We study how the change in the three model parameters affects  $P^s_0(k)$
	and $P^s_2(k)$.

	The top, middle and bottom panels of Figure~\ref{fig:sim-monopole} respectively show the
	variation of $P^s_0(k)$ with respect to the parameters $\alpha$, $h_f$ and $\sigma_v$
	of the LP method. The HC and the HP methods are also shown for comparison, and 
	all the results have been normalized using $[P^s_0(k)]_{\rm HC}$ of the HC method.
	We observe that at all redshifts, 
	$P^s_0(k)$ for all the methods are indistinguishable at small $k$ ($k < 0.4\,{\rm Mpc}^{-1}$). 
	This indicates that the large scale coherent flows are same for the different 
	methods considered here. However, the LP and the HP methods differ from the HC method
	at large $k$ (small scales) where we observe suppression in $P^s_0(k)$ due to the FoG effect, 
	and this suppression is maximum for the HP method and it is somewhat less for the LP method. 
	The $k$ value at which we observe this difference decreases with decreasing redshift, and for 
	the LP method at a fixed redshift, this $k$ value changes slightly with a change in 
	the model parameters. For both the LP and the HP methods, the FoG suppression 
	increases with increasing $k$ and decreasing $z$. 
	Considering the variation in $\alpha$ with respect to the reference model, 
	we see that the suppression is maximum at the reference value
	$\alpha=0$ which corresponds to a symmetric line profile where both the horns
	are identical (Figure~\ref{fig:model-profiles}). The suppression reduces if 
	$\alpha$ is increased which is equivalent to the increase in asymmetry in the line profile. 
	Considering the variation with $h_f$, we see that the suppression is maximum at the
	reference value $h_f=0.1$ for which a very small fraction of \HI is contained in the
	core. The suppression reduces with increasing $h_f$ that is as the $\HI$ fraction in the
	core increases in the expense of the disk. 
	Considering the variation with $\sigma_v$, we see that the suppression is maximum for 
	$\sigma_v=20 \,{\rm km\,s}^{-1}$ which corresponds to large random motions.
	The suppression decreases with decreasing $\sigma_v$ or as the random 
	motions are reduced. The change in the $P^s_0(k)$ due to the variation 
	in the three parameters of the LP method	is small in comparison to 
	the difference between the HC and the HP methods. The results of the 
	LP method are closer to the HP method than that of the HC method.

	%----------------------------------- Figure ---------------------------		
	\begin{figure*}
		
		\psfrag{sigmap}[c][c][1.5][0]{$\sigma_p$}
		\psfrag{z}[c][c][1.5][0]{$z$}
		
		\psfrag{B1}[c][c][1.5][0]{B1}
		\psfrag{B2}[c][c][1.5][0]{B2}
		\psfrag{B3}[c][c][1.5][0]{B3}

		\psfrag{al=0.0}[c][c][1][0]{$\,\alpha=0\qquad\quad$}
		\psfrag{al=0.4}[c][c][1][0]{$0.4$}
		\psfrag{al=0.8}[c][c][1][0]{$0.8$}
		\psfrag{al=1.0}[c][c][1][0]{$1.0$}
		
		\psfrag{hf=0.1}[c][c][1][0]{$h_f=0.1\qquad\,\:$}
		\psfrag{hf=0.4}[c][c][1][0]{$0.4$}
		\psfrag{hf=0.8}[c][c][1][0]{$0.8$}
		\psfrag{hf=1.0}[c][c][1][0]{$1.0$}
		
		\psfrag{sv=10.0}[c][c][1][0]{$\sigma_v=10 \,{\rm km\,s}^{-1}\qquad\quad$}
		\psfrag{sv=5.0}[c][c][1][0]{$5 \,{\rm km\,s}^{-1}\,\,\,$}
		\psfrag{sv=15.0}[c][c][1][0]{$15 \,{\rm km\,s}^{-1}\,\,\,$}
		\psfrag{sv=20.0}[c][c][1][0]{$20 \,{\rm km\,s}^{-1}\,\,\,$}
		
		\psfrag{hf=0.1 sv=10.0}[c][c][1][0]{$h_f=0.1,\,\sigma_v=10 \,{\rm km\,s}^{-1}\quad$}
		\psfrag{al=0.0 sv=10.0}[c][c][1][0]{$\alpha=0,\,\sigma_v=10 \,{\rm km\,s}^{-1}\quad$}
		\psfrag{al=0.0 hf=0.1}[c][c][1][0]{$\alpha=0,\,h_f=0.1$}

		\psfrag{HP}[c][c][1][0]{HP}
		\psfrag{HC}[c][c][1][0]{HC}
		\psfrag{LT}[c][c][0.9][0]{Linear Theory$\qquad  \qquad \quad$}
		\psfrag{HI multipole ratio}[c][c][1][0]{$P^s_{2}/P^s_{0}$}
		\psfrag{kmode}[c][c][1][0]{$k\, {\rm Mpc}^{-1}$}
		
		\psfrag{z=1}[c][c][1.3][0]{$z=1$}
		\psfrag{z=2}[c][c][1.3][0]{$z=2$}
		\psfrag{z=3}[c][c][1.3][0]{$z=3$}
		\psfrag{z=4}[c][c][1.3][0]{$z=4$}
		\psfrag{z=5}[c][c][1.3][0]{$z=5$}
		\psfrag{z=6}[c][c][1.3][0]{$z=6$}

		\centering
		\includegraphics[width=0.5\textwidth,angle=-90]{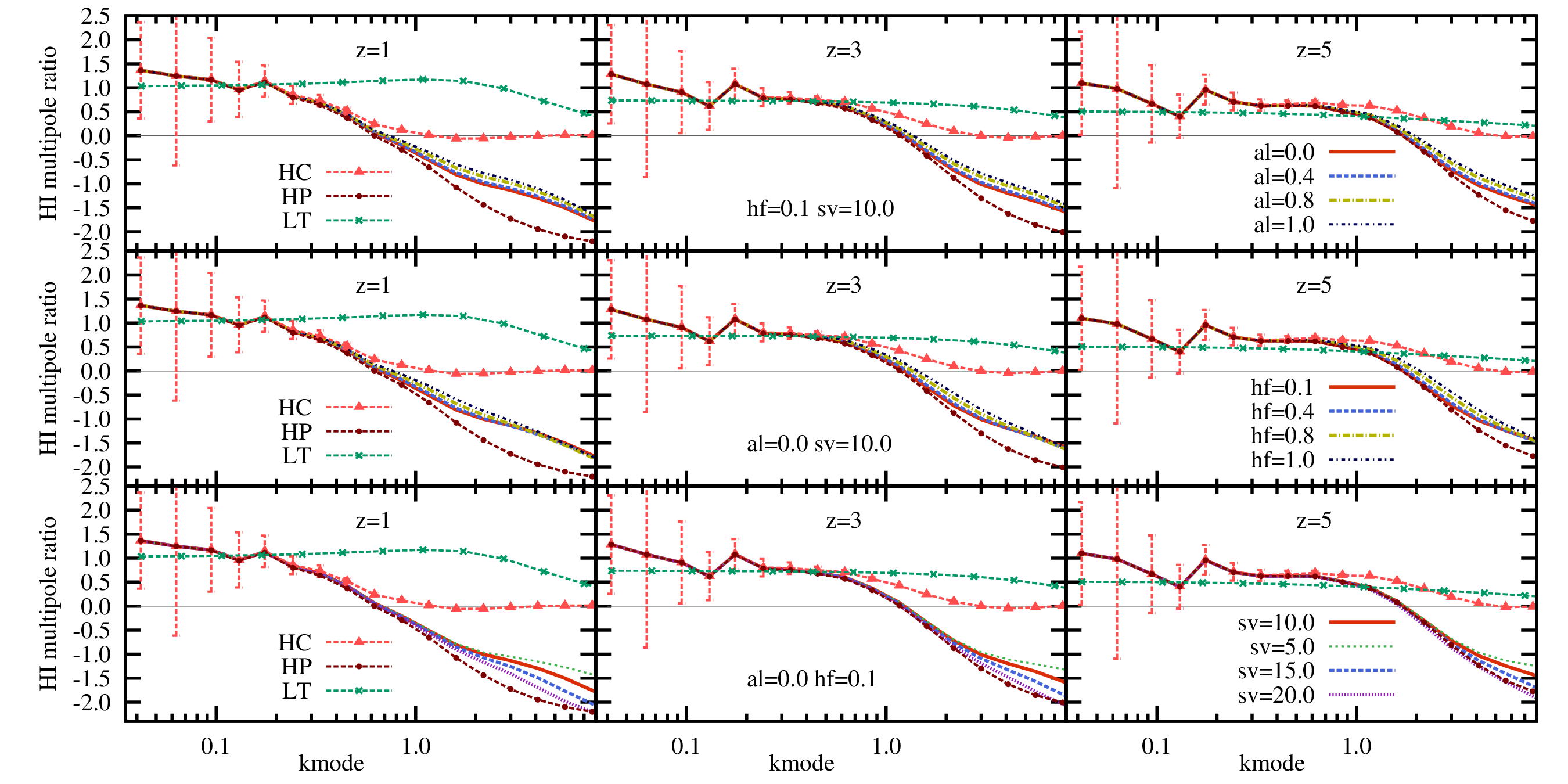}

		\caption{This shows the variation of the  ratio 
			$P^s_2(k)/P^s_0(k)$ with respect to 
			the parameters $\alpha$ (upper panels), $h_f$ (middle panels) 
			and $\sigma_v$ (lower panels) of the LP method at three different redshifts.
			The variations in $\alpha$, $h_f$ and $\sigma_v$ are with respect to the reference
			values $\alpha=0, h_f=0.1, \sigma_v=10 \,{\rm km\,s}^{-1}$. The results for 
			the HC and the HP methods along with the linear theory predictions 
			(Equation~\ref{eq:LT_prediction}) are also shown for comparison. Considering 
			the HC method, the $1-\sigma$ spread determined from the five independent
			realizations of the simulations is also shown.}
		
		\label{fig:sim-ratio}
		
	\end{figure*}
	%----------------------------------- Figure ---------------------------		

	We next study the quadrupole $P^s_2(k)$. Here we have considered the ratio $P^s_2(k)/P^s_0(k)$
	which can be determined directly from observations and which  contains important cosmological information.
	On large scales where the coherent flows dominate the RSD, linear theory predicts this ratio to be 
	\citep{hamilton-98-rsd-review}
	%------------------------------
	\begin{equation}
	\frac{P^s_2(k)}{P^s_0(k)}=\frac{(4/3)\beta + (4/7)\beta^2}{1+ (2/3)\beta + (1/5) \beta^2}\,,
	\label{eq:LT_prediction}
	\end{equation}
	%------------------------------
	where $\beta=f(\Omega_m)/b$ is the redshift distortion parameter. 
	Here the Kaiser enhancement ensures that $P^s_2(k)$ is positive and, therefore
	the ratio  also is predicted to be positive. A measurement of this ratio at large scales, where linear theory is expected to be valid, holds 
	the promise of determining $\beta$. Provided we independently know the \HI bias $b$, 
	the ratio can further be used to determine the growth rate of density 
	perturbations $f(\Omega_m)$ and also the cosmological matter density parameter $\Omega_m$.

	The top, middle and bottom panels of Figure~\ref{fig:sim-ratio} respectively 
	show the variation of the ratio $P^s_2(k)/P^s_0(k)$ with respect to the parameters 
	$\alpha$, $h_f$ and $\sigma_v$ of the LP method. The linear theory predictions 
	(Equation~\ref{eq:LT_prediction}) for a scale dependent \HI bias b(k) 
	(Paper I) are also shown for reference. We see that at all redshifts the ratio $P^s_2(k)/P^s_0(k)$ determined from  the different methods all overlap  at small $k$ ($k < 0.4\,{\rm Mpc}^{-1}$). The RSD here is dominated by the coherent flows which is the same for all the methods. The ratio $P^s_2(k)/P^s_0(k)$  here is nearly scale independent and is consistent with the linear theory predictions 
	(Equation~\ref{eq:LT_prediction}). 
	The results for the three different methods, however, differ at large $k$ 
	where the FoG effect dominates. We see that the ratio is minimum for the HP method and 
	maximum for the HC method. The results for the LP method lie in between the HC and
	the HP methods, and they are  closer to the HP method. The $k$ value at which the 
	three methods start to differ from each other decreases with decreasing redshift.
	At a fixed redshift, we observe that  for all the methods $P^s_2(k)/P^s_0(k)$
	crosses zero at a particular $k$ value which we denote as $k_0$, and 
	the ratio becomes negative at $k>k_0$. This $k$ value of zero crossing ($k_0$) corresponds
	to a spatial length scale above which ({\it i.e.} at $k<k_0$) the Kaiser effect dominates and $P^s_2(k)$
	is positive, and below this length scale ({\it i.e.} at $k>k_0$) the FoG effect dominates which makes
	$P^s_2(k)$ negative. We see that $k_0$ is minimum for the HP method and 
	maximum for the HC method, and for the LP method $k_0$ changes slightly with a 
	change in the model parameters. The change in the ratio $P^s_2(k)/P^s_0(k)$ 
	due to the variation in the three parameters of the LP method is small in 
	comparison to the difference between the HC and the HP methods. 
	Considering the LP method at large $k$, we see that the ratio increases with
	increasing $\alpha$ and $h_f$, and decreases with increasing $\sigma_v$.

	%----------------------------------- Figure ---------------------------		
	\begin{figure}
		
		\psfrag{sigmap}[c][c][1.5][0]{$\sigma_p\,{\rm Mpc}$}
		\psfrag{z}[c][c][1.5][0]{$z$}

		\psfrag{HP}[c][c][1.4][0]{HP}
		\psfrag{HC}[c][c][1.4][0]{HC}
		\psfrag{LP}[c][c][1.4][0]{LP}
		
		\centering
		\includegraphics[width=0.4\textwidth,angle=-90]{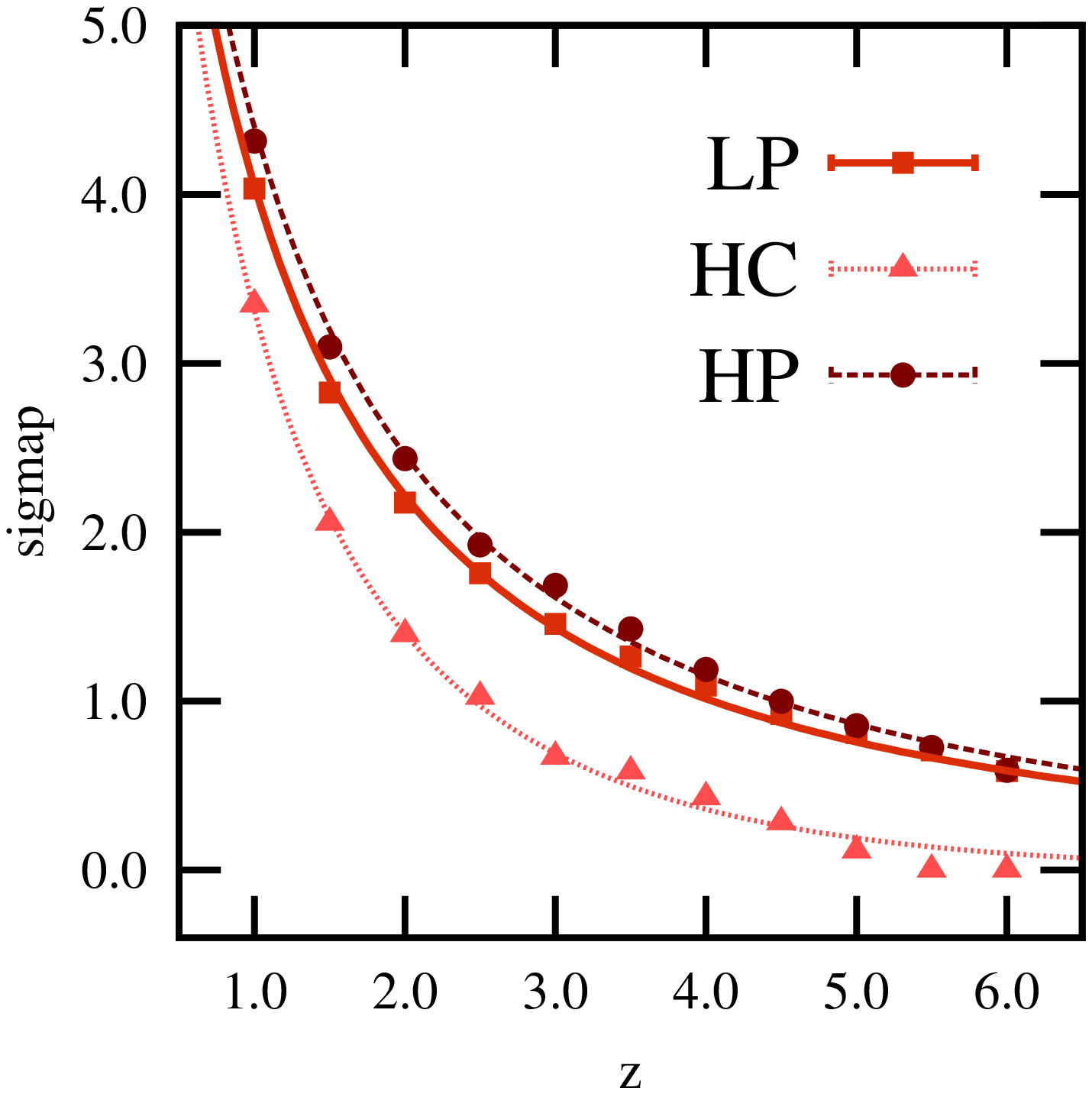}

		\caption{The circles, the squares and the triangles respectively show 
			the best-fitting values of $\sigma_p$ at different redshifts for 
			the HP, the reference LP $(\alpha=0, h_f=0.1,\sigma_v=10\,{\rm km\,s}^{-1})$ 
			and the HC methods. The predictions of Equation~\ref{eq:sigmap_model} 
			are shown with  different line styles.}
		\label{fig:best-fit-sigmap}
		
	\end{figure}
	%----------------------------------- Figure ---------------------------		

	In Paper II, we have considered a model for $P^s_{\HI}(\kpe, \kpa)$ of the form 
	\citep{peacock92,park-vogeley94,peacock-dodds94,ballinger-peacock-heavens96}
	%--------------------
	\begin{equation}
	P^s_{\HI}(\kpe,\kpa)=b^2(k) \left(1+\beta \mu^2 \right)^2 P(k) D_{FoG}(\kpa,\sigma_p) \,.
	\label{eq:model-pk}
	\end{equation} 
	%--------------------
	Here, $P(k)$ is the real space dark matter power spectrum, $b(k)$ is the \HI bias 
	and $\mu=\kpa/k$ is the cosine of the angle between the $\bf{k}$ vector and 
	the line of sight. The factor $\left(1+\beta \mu^2 \right)^2$  incorporates the 
	Kaiser enhancement and the damping profile 
	$D_{FoG}(\kpa,\sigma_p)= \left( 1+ \frac{1}{2} \kpa^2 \sigma_p^2 \right)^{-1}$,
	which we have considered to be a Lorentzian 
	\citep{hatton-cole99, seljak01-RSD-halo-model, white01-RSD-halo-model},
	incorporates the FoG suppression. Here, $\sigma_p$ is the pairwise 
	velocity dispersion which is in units of comoving Mpc, one can equivalently use 
	$[\sigma_p a H(a)]$ in units of ${\rm km\,s}^{-1}$. Using the real space
	simulations we can completely calculate the Kaiser enhancement term.
	The model therefore has only one unknown parameter $\sigma_p$.

	In Paper I, we have considered the possibility of a complex \HI bias $\tilde{b}(k)$ 
	which is equivalent to a stochasticity parameter $r$. In Paper II, we have included
	this possibility in the Kaiser enhancement term. Further, in addition to 
	the Lorentzian profile considered here, in Paper II we have also considered the Gaussian and
	the Lorentzian Squared damping profiles for the FoG effect. However, our earlier work shows
	that it is adequate to restrict the analysis to $r=1$ (no stochasticity) and only consider the
	Lorentzian profile for the FoG damping.

	The model in Equation~\ref{eq:model-pk} has a single free parameter $\sigma_p$.
	Using this model, we perform $\chi^2$ minimization with respect to $\sigma_p$ 
	to determine the best value that fits the simulated $P^s_{\HI}(\kpe,\kpa)$.
	While the model works well at small $k$, it tends to deviate significantly from 
	the simulations at large $k$. Further, the deviation is seen to increase 
	at lower redshifts. The deviation at large $k$ greatly influences the fitting 
	procedure and we find that it is better to exclude the large $k$ values for 
	fitting the simulated $P^s_{\HI}(\kpe,\kpa)$. Following Paper II, we have restricted the
	values of $(\kpe,\kpa)$ to be within $\{0.9\,,1.1\,,1.5\,,1.7\,,2.0\}\,{\rm Mpc}^{-1}$ 
	for the redshifts $\{ 1\,,1.5\,,2\,,2.5\,,\geq3 \}$. We determine the goodness 
	of fit on the basis of the reduced chi-square $\chi^2/N$, where $N$ is the
	degree of freedom.	We find that the model yields $\chi^2/N \lesssim 1\,$ at all redshifts.

	Figure~\ref{fig:best-fit-sigmap} shows the variation of the best fit $\sigma_p(z)$ values with $z$. 
	We see that at all redshifts $\sigma_p$ is largest for the HP method, followed by LP and then HC. 
	The $\sigma_p$ value quantifies the relative strength of the FoG effect. The values of $\sigma_p$ 
	as well as the behaviour of the monopole  and the quadrupole (Figures~\ref{fig:sim-monopole} and~\ref{fig:sim-ratio}) all clearly illustrate that the FoG effect is strongest for the HP,  followed by the  LP and then the HC method. For all the methods we see that  $\sigma_p$  decreases  with increasing $z$.  This decline is fastest for the HC method where $\sigma_p$ flattens off at a value  consistent with zero  at $z>5\,$. For the LP and the HP methods the $\sigma_p$ values are quite a bit larger than those for the HC method,  and they decline less steeply with $z$. For these two methods, $\sigma_p$ has a non-zero value even at our largest redshift $z=6\,$.

	Following Paper II, we model the $z$ dependence	of $\sigma_p$ using 
	%--------------------------------------
	\begin{equation}
	\sigma_p(z)=\sigma_p(0) \left( 1+z \right)^{-m} \exp \left[-\left( \frac{z}{z_p} \right)^2 \right]\,,
	\label{eq:sigmap_model}
	\end{equation}
	%--------------------------------------
	where $\sigma_p(0)$, $m$ and $z_p$ are three fitting parameters. Here, $\sigma_p(0)$ represents
	the extrapolated $\sigma_p$ value at $z=0$, $m$ is the slope at low $z$ and 
	$\exp \left[-\left( z/z_p \right)^2 \right]$ incorporates the flattening of $\sigma_p$
	at high $z$.  Considering the (HC, LP, HP) methods in Figure~\ref{fig:best-fit-sigmap} we obtain 
	$\sigma_p(0)=(12.6,11.3,11.7)\,{\rm Mpc}$, $m=(1.88,1.47,1.41)$
	and $z_p=(5.5,20,17.5)$ from a rough fitting of $\sigma_p(z)$ as a function of $z$. 
	The  larger slope ($m=1.88$) for the HC method reflects the  fact that for 
	this method  $\sigma_p$ declines faster with increasing $z$ as compared to 
	the LP and HP methods. We also see that the $\exp \left[-\left( z/z_p \right)^2 \right]$ 
	term is only relevant for the HC method as $\sigma_p$ does not become zero within $z \leqslant 6$ 
	for the other two methods.  It may be noted that the  values of $\sigma_p(0),m,z_p$ quoted here for the HC and HP methods differ	from those in Paper II which had implemented  	a different \HI mass assignment scheme  for the haloes.

	%----------------------------------- Figure ---------------------------		
	\begin{figure*}
		
		\psfrag{sigmap}[c][c][1.5][0]{$\Delta \sigma_p\,{\rm Mpc}$}
		\psfrag{z}[c][c][1.5][0]{$z$}
		
		\psfrag{al=0.0}[c][c][1][0]{$\,\alpha=0.0$}
		\psfrag{al=1.0}[c][c][1][0]{$\,\alpha=1.0$}
		
		\psfrag{hf=0.1}[c][c][1][0]{$\,h_f=0.1$}
		\psfrag{hf=1.0}[c][c][1][0]{$\,h_f=1.0$}
		
		\psfrag{sv=20.0}[c][c][0.9][0]{$\,\,\qquad\sigma_v=20 \,{\rm km\,s}^{-1}$}
		\psfrag{sv=5.0}[c][c][0.9][0]{$\,\,\qquad\sigma_v=5 \,{\rm km\,s}^{-1}$}
		
		\psfrag{HP}[c][c][1.1][0]{HP}
		
		\centering
		\includegraphics[width=0.33\textwidth,angle=-90]{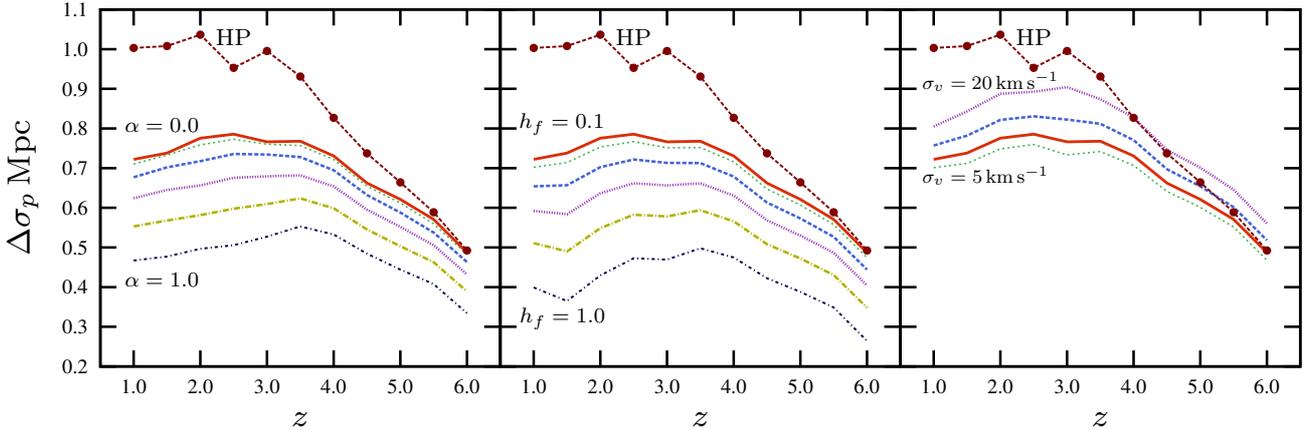}
		
		\caption{This shows $\Delta \sigma_p(z)$ the difference in $\sigma_p(z)$ 
			relative to the HC method. Considering the LP method, the left, 
			central and right panels respectively show the variation with  
			the parameters $\alpha\, (=0,0.2,0.4,...,1)$, $h_f\, (=0.1,0.2,0.4,...,1)$ 
			and $\sigma_v \, (=5,10,15,20 \, {\rm km\,s}^{-1})$.	
			The results for the HP method (dashed line with circles) 
			are also shown for comparison.}
		
		\label{fig:sigmap-diff-wrt-HC}
		
	\end{figure*}
	%----------------------------------- Figure ---------------------------		

	Considering the overall variation  in $\sigma_p(z)$  
	between the different methods, Figure~\ref{fig:sigmap-diff-wrt-HC} shows  
	$\Delta \sigma_p(z)$ the difference relative to the HC method. We see that 
	$\Delta \sigma_p$  is largest for the HP method where it increases from 
	$\Delta \sigma_p\approx 0.5 \, {\rm Mpc}$ at $z=6$ and saturates at  
	$\Delta \sigma_p \approx 1 \, {\rm Mpc}$ at $z \leqslant 3\,$. 
	Considering the LP method, we see that the reference model shows a behaviour 
	similar  to the HP method, however  the increase in $\Delta \sigma_p$ 
	with decreasing $z$ is slower and the peak value $(0.7-0.8 \, {\rm Mpc})$ 
	is smaller. Considering the variation with $\alpha$ (left panel), we see that 
	$\Delta \sigma_p$ shows an overall decline as $\alpha$ is varied from $0$ 
	(reference model) to $1\,$. Note that 
	$\alpha=0$ corresponds to a symmetric line profile where both the horns are 
	identical (Figure~\ref{fig:model-profiles}) and the asymmetry between the two 
	horns is maximum for  $\alpha=1\,$. Likewise, the values of $\Delta \sigma_p$  
	also fall if $h_f$ is varied (central panel) from $0.1$ to $1\,$. Here $h_f$ 
	quantifies the fraction of the total \HI contained in the core. The values of  
	$\Delta \sigma_p$  increase as $\sigma_v$ is increased (right panel) 
	from $5\,{\rm km\,s}^{-1}$  to $20\,{\rm km\,s}^{-1}$. Here $\sigma_v$   
	quantifies the \HI random motions (thermal and turbulent).  
	Considering all these results together, two interesting features emerge from this analysis. 
	First,  the value of $\sigma_p(z)$ (or equivalently the FoG effect) is significantly 
	larger than that predicted by the HC method which ignores the motion of the \HI 
	within the galaxies. Second, we see that   the deviations from the HC method 
	($\Delta \sigma_p(z)$) increases in the initial stages of evolution (high $z$) 
	and then nearly saturates at $z \leqslant 3\,$. This indicates that contribution from 
	the \HI motions within  the galaxies saturates at $z \approx 3$ and the 
	subsequent increase in $\sigma_p(z)$  seen in Figure~\ref{fig:best-fit-sigmap} 
	is due to the growth of the peculiar velocities of the haloes. We note that 
	$\Delta \sigma_p(z)$ declines slightly at $z <3$, we however currently do not 
	have an understanding of this bahaviour.

	%----------------------------------- Figure ---------------------------		
	\begin{figure*}
		
		\psfrag{HI monopole}[c][c][1.3][0]{$P^s_0$}
		\psfrag{HI quadrupole}[c][c][1.3][0]{$P^s_2$}
		\psfrag{HI multipole ratio}[c][c][1.3][0]{$P^s_2/P^s_0$}		
		
		\psfrag{kmode}[c][c][1.3][0]{$\qquad k\, {\rm Mpc}^{-1}$}
		
		\psfrag{LP}[c][c][1.2][0]{LP}		
		\psfrag{HP}[c][c][1.2][0]{HP}
		\psfrag{HC}[c][c][1.2][0]{HC}

		\psfrag{z=1}[c][c][1.5][0]{$\quad z=1$}
		\psfrag{z=3}[c][c][1.5][0]{$\quad z=3$}
		\psfrag{z=5}[c][c][1.5][0]{$\quad z=5$}

		\centering
		\includegraphics[width=0.5\textwidth,angle=-90]{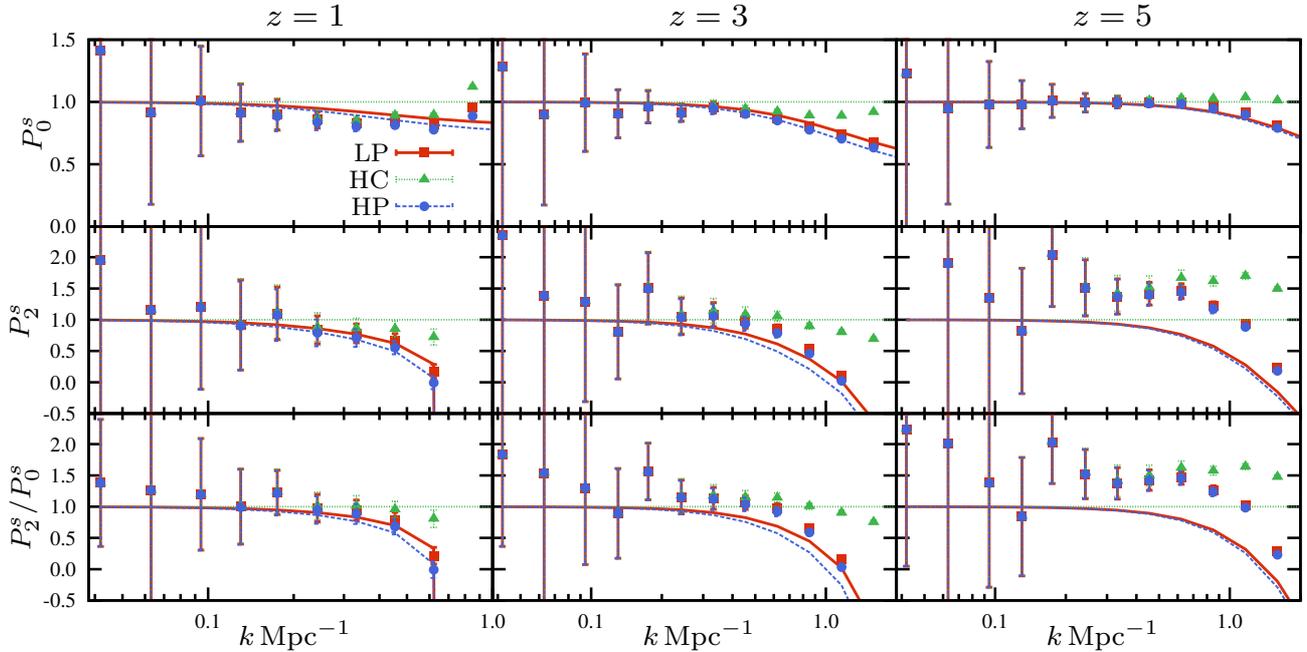}

		\caption{Considering  $P^s_{0}$, $P^s_{2}$  and the ratio $P^s_{2}/P^s_{0}$ 
				in the top, middle and bottom rows respectively, the data points with 
				$1-\sigma$ error bars show the  results from simulations with the HC, 
				HP and reference LP methods using different point styles as indicated 
				in the figure. The lines show the corresponding model predictions 
				(Equation~\ref{eq:model-pk}). The results are all  
				normalized by  the model prediction for the HC method.  The left, 
				middle and right columns  correspond to $z=1,3$ and $5$ respectively.}	
		\label{fig:best-fit-model}

	\end{figure*}
	%----------------------------------- Figure ---------------------------		

	Considering the model in Equation~\ref{eq:model-pk}, we now discuss  
	the $k$ range over which its  predictions match the monopole $P^s_0(k)$, 
	the quadrupole $P^s_2(k)$ and the ratio $P^s_2(k)/P^s_0(k)$ obtained 
	from the simulations. Figure~\ref{fig:best-fit-model} shows the simulated 
	values along with the model predictions	of $P^s_{0}(k)$ (top row), $P^s_{2}(k)$ 
	(middle row) and $P^s_{2}(k)/P^s_{0}(k)$ (bottom row).	
	We find that the results for the LP method are 
	very similar to those we have obtained earlier for the HC and HP methods
	in Paper II. While the model predictions match  the simulated $P^s_0(k)$ 
	at all redshifts for almost the entire  
	$k$ range  used for the fitting,  the  model predictions match the 
	simulated $P^s_2(k)$ (or the ratio $P^s_2(k)/P^s_0(k)$) for a  
	limited $k$ range. At $z \leqslant 3$, the model matches the simulated 
	$P^s_2(k)$ up to $k<0.5\,{\rm Mpc}^{-1}$, whereas this $k$ range shrinks to
	$k<0.2\,{\rm Mpc}^{-1}$ at $z \geqslant 4$ and the model underpredicts 
	$P^s_2(k)$ beyond this $k$ range.

	\section{Summary and Discussion}
	\label{sec:summary-discussion}
	
	The post-reionization \HI 21-cm signal is a very promising probe of the 
	large scale structures in the Universe. It will, in principle, be possible 
	to measure the growth rate of linear perturbations $f(\Omega_m)$  provided one 
	is able to correctly model the redshift space distortion in the 21-cm signal
	\citep{castorina19-growth}. This is also relevant for predicting the 21-cm 
	power spectrum signal expected at different instruments, and for correctly 
	interpreting the signal once a detection is made.  Redshift space distortions 
	has been extensively studied in the context of galaxy surveys \citep{hamilton-98-rsd-review,lahav04-rsd-review}. Here it is necessary  
	to appreciate an important difference between galaxy surveys and 21-cm intensity mapping. 
	The individual galaxies are the fundamental elements in galaxy redshift surveys. 
	The surveys are only sensitive to the peculiar motion of the
	individual galaxies and motions inside the galaxies are of no consequence for redshift space distortions. In contrast, 21-cm intensity mapping does not see the
	individual sources but only sees the collective radiation from all the
	sources. The signal is sensitive to all motions, be it that of the galaxies
	which contain the \HI or the motion of the \HI inside the galaxies.
	The intensity mapping experiments also have high enough 
	frequency resolution ($< 100$ kHz or $< 20\,{\rm km\,s}^{-1}$) to resolve
	the velocity structure of \HI within the galaxies, and 
	the RSD effect here is quite different from that in galaxy surveys.

	Low redshift observations indicate that the \HI is predominantly 
	contained in the rotating  spiral galaxies for which  we have modelled 
	the 21-cm emission  using a double horned line profile 
	(Figure~\ref{fig:model-profiles}). The LP method (Section~\ref{subsec:simulating-LP}) 
	incorporates  the line profiles of the individual galaxies into the simulated 
	\HI 21-cm signal. In this picture the horns arise from the \HI disk whereas 
	the rest of the \HI is contained in the core. As discussed in Section~\ref{subsec:simulating-LP}, 	
	the parameter $h_f$ quantifies the fraction of the \HI in the core whereas the parameter  
	$\alpha$  ( $-1 \leqslant \alpha \leqslant 1$)  quantifies the asymmetry between the 
	two horns which  are completely symmetric for $\alpha=0\,$.  In addition to this, 
	we also include a random velocity component  through a third parameter $\sigma_v$ 
	which accounts for both  thermal and turbulent motions. 
	For comparison, we have also considered simulations using the HC method where all 
	the \HI inside a galaxy is assumed to move with the same peculiar velocity as the 
	halo which hosts the galaxy, and  the HP method where the \HI mass elements within 
	a galaxy are all assigned different  peculiar velocities corresponding to  
	the individual dark matter simulation particles that constitute the host halo.

	We find that  the redshift space \HI power spectrum $P^s_{\HI}(\kpe,\kpa)$ (Figure~\ref{fig:2dpowerspectrum}) for all the methods match at small $k$ 
	where the results are consistent with the Kaiser enhancement  \citep{kaiser87} 
	arising from coherent flows. The results differ at large $k$ where we have the FoG  
	suppression. This suppression is most pronounced for the HP method followed by 
	the LP and HC methods. The HC method does not show any FoG suppression at $z \ge 5\,$. 
	We have modelled the FoG suppression using a Lorentzian damping profile 
	(Equation~\ref{eq:model-pk}) which has a free parameter $\sigma_p$ the pairwise 
	velocity dispersion which quantifies the relative strength of the FoG effect. 
	The value of $\sigma_p(z)$ has been determined by fitting the simulated 
	$P^s_{\HI}(\kpe,\kpa)$.	We see that at all redshifts the best fit $\sigma_p(z)$
	value is largest for the HP method, followed by LP and then HC 
	(Figure~\ref{fig:best-fit-sigmap}). Considering the LP method with the reference 
	parameter values $(\alpha,h_f,\sigma_v)=(0,0.1, 10 \, {\rm km\,s}^{-1})$,  we find 
	that the $\sigma_p(z)$ values are closer to the HP method than those of the HC method. 
	For all the methods we find that  $\sigma_p(z)$  decreases  with increasing $z$ 
	approximately as $(1+z)^{-m}$. This decline is fastest for the HC method where 
	we have $m=1.88$ and the $\sigma_p(z)$ values are consistent with zero for $z \geq 5\,$. 
	This decline is relatively slower for the other two methods with $m=1.47$ and $1.41$ 
	for the LP and the HP methods respectively, and $\sigma_p(z)$ has a finite non-zero 
	value for the entire $z$ range considered here.

	The HC method incorporates only the peculiar velocity of the entire galaxy 
	(assumed to be the same as that of the host halo) and ignores the motion of 
	the \HI inside the galaxy. We have the smallest FoG suppression and $\sigma_p(z)$ 
	values in this case. The LP method incorporates the rotational and random motions 
	within the galaxy through the line profile which results in an enhancement in the 
	FoG suppression and $\sigma_p(z)$ values. The HP method incorporates the extreme 
	situation where the \HI mass elements have the same velocity dispersion as the 
	dark matter simulation particles that constitute the halo, and we see the maximum FoG 
	suppression and $\sigma_p(z)$ values in this case. We have considered the
	quantity $\Delta \sigma_p(z)$ (Figure~\ref{fig:sigmap-diff-wrt-HC}) which is 
	the excess in $\sigma_p(z)$  over the HC method to quantify the extra FoG suppression 
	due to the motions within the galaxy.	We find that $\Delta \sigma_p(z)$ 	is largest 
	for the HP method where it has a value	$\Delta \sigma_p(z)\approx 0.5 \, {\rm Mpc}$ 
	at $z=6$, 	and increases with decreasing $z$  and saturates at  
	$\Delta \sigma_p(z) \approx 1 \, {\rm Mpc}$ at $z \leqslant 3\,$. 
	The reference LP method shows a behaviour similar  to the HP method, however  
	the increase in $\Delta \sigma_p(z)$ 
	with decreasing $z$ is slower and the peak value $(0.7-0.8 \, {\rm Mpc})$ 
	is smaller. Considering the LP method, we see that  $\Delta \sigma_p(z)$ 
	decreases if either the asymmetry between the two horns ($\alpha$) increases or 
	the \HI fraction in the core ($h_f$) increases whereas  $\Delta \sigma_p(z)$  
	increases with $\sigma_v$.  Considering all these results together, two interesting 
	features emerge from this analysis. First,  the value of $\sigma_p(z)$ 
	(or equivalently the FoG effect) is significantly larger than that predicted by 
	the HC method which ignores the motion of the \HI within the galaxies. Second, 
	we see that   the deviations from the HC method ($\Delta \sigma_p(z)$) increases 
	in the initial stages of evolution (high $z$) and then saturates at $z \leqslant 3\,$. 
	This indicates that contribution from the \HI motions within  the galaxies saturates 
	at $z \approx 3$ and the 	subsequent increase in $\sigma_p(z)$  seen in 
	Figure~\ref{fig:best-fit-sigmap} 
	is due to the growth of the peculiar velocities of the haloes. As $\sigma_p(z)$ 
	at high redshifts arises   mainly  from the \HI motions within the galaxies (or haloes), 
	a measurement  here will essentially provide a handle on the line profiles  of high 
	redshift galaxies. Comparing our results with some of the recent works, we note that  
	the values of $\sigma_p(z)$ inferred in \citet{navarro-ingredients-for-IM2018} and  
	\citet{ando-rsd-2018} are consistent with our HC method which indicates that these 
	simulations possibly underestimate the motion of the \HI within the galaxies.

	The main outcome of our entire  analysis is that the motion of \HI within the 
	galaxies makes a significant contribution to the redshift space distortion and 
	this will be manifested as an enhancement in the Finger of God effect.  In this 
	work we have proposed a simple technique to incorporate this in 21-cm simulations
	through a galaxy line profile. For the present work we have made a simplifying 
	assumption that all the galaxies in	a simulation have the same $\alpha$, $h_f$ 
	and $\sigma_v$ values. In reality we expect the parameter  values to vary from 
	galaxy to galaxy, though the variation in $\sigma_p$ with a change in these 
	parameters is not  very large (Figure \ref{fig:sigmap-diff-wrt-HC}) we plan 
	to address this in future work. In addition to this, our present analysis 
	assumes that the high mass haloes  ($M_h >  M_{\rm max}$) do not contain \HI. 
	In reality these haloes could  contain multiple galaxies, central and satellites,  
	which may  make a non-neglible  contribution to the total \HI budget and also 
	lead to an enhanced FoG. We propose to  consider this possibility in future work.
	
	\section*{Acknowledgements}

	We would like to thank the reviewer for the  careful and thorough reading
	of this manuscript and for the thoughtful comments and constructive 
	suggestions which help to improve the quality of this manuscript.	
	
%------------------------------------- Bibliography--------------------------------------------------
%	\bibliography{reference}

%-------------- End Bibliography ---------------------------------------------------------------------	
	
	% Don't change these lines
	\bsp	% typesetting comment
	\label{lastpage}
\end{document}